\documentclass{aastex631}
\usepackage{natbib}
\usepackage{amsmath}
\usepackage{verbatim}
\usepackage{hyperref}
\usepackage{float}
\usepackage{graphicx}
\usepackage{mathtools}
\usepackage{empheq}
\usepackage{appendix}
\usepackage{microtype}
\usepackage{commath}
\usepackage{bold-extra}
\usepackage{comment}
\usepackage{rotating}

\DeclareGraphicsExtensions{.pdf,.png,.jpg}
\DeclareUnicodeCharacter{2223}{$|$}

\newcommand{\kmsmpc}{\hbox{$ \, \rm km\, s^{-1} \, Mpc^{-1}$}}

\newcommand{\bq}{\begin{equation}} 
\newcommand{\eq}{\end{equation}} 
 
\newcommand{\beq}{\begin{equation}}
\newcommand{\eeq}{\end{equation}}
\newcommand{\beqa}{\begin{eqnarray}}
\newcommand{\eeqa}{\end{eqnarray}}

\newcommand{\PL}{$P$--$L$\ }

\defcitealias{Riess:2023}{R23}

\defcitealias{Riess:2022}{R22}

\newcommand{\MB}{-19.28}
\newcommand{\MBh}{72.5}

\received{August 23, 2024}

\shorttitle{Context for Comparing {\it JWST} and {\it HST}}
\shortauthors{Riess et al.}

\begin{document}

\title{{\it JWST} Validates {\it HST} Distance Measurements: \\   Selection of Supernova Subsample Explains Differences in {\it JWST} Estimates of Local H$_0$}

\author[0000-0002-6124-1196]{Adam G.~Riess}
\affiliation{Space Telescope Science Institute, 3700 San Martin Drive, Baltimore, MD 21218, USA}
\affiliation{Department of Physics and Astronomy, Johns Hopkins University, Baltimore, MD 21218, USA}

\author[0000-0002-4934-5849]{Dan Scolnic}
\affiliation{Department of Physics, Duke University, Durham, NC 27708, USA}

\author[0000-0002-5259-2314]{Gagandeep S. Anand}
\affiliation{Space Telescope Science Institute, 3700 San Martin Drive, Baltimore, MD 21218, USA}

\author[0000-0003-3889-7709]{Louise Breuval}
\affiliation{Department of Physics and Astronomy, Johns Hopkins University, Baltimore, MD 21218, USA}

\author[0000-0000-0000-0000]{Stefano Casertano}
\affiliation{Space Telescope Science Institute, 3700 San Martin Drive, Baltimore, MD 21218, USA}

\author[0000-0002-1775-4859]{Lucas M.~Macri}
\affiliation{NSF NOIRLab, 950 N Cherry Ave, Tucson, AZ 85719, USA}

\author[0000-0002-8623-1082]{Siyang Li}
\affiliation{Department of Physics and Astronomy, Johns Hopkins University, Baltimore, MD 21218, USA}

\author[0000-0001-9420-6525]{Wenlong Yuan}
\affiliation{Department of Physics and Astronomy, Johns Hopkins University, Baltimore, MD 21218, USA}

\author[0000-0001-6169-8586]{Caroline D. Huang}
\affiliation{Center for Astrophysics ∣ Harvard \& Smithsonian, 60 Garden Street, Cambridge, MA 02138, USA}

\author[0000-0001-8738-6011]{Saurabh Jha}
\affiliation{Department of Physics and Astronomy, Rutgers, The State University of New Jersey, 136 Frelinghuysen Road, Piscataway, NJ 08854}

\author[0000-0002-8342-3804]{Yukei S. Murakami}
\affiliation{Department of Physics and Astronomy, Johns Hopkins University, Baltimore, MD 21218, USA}

\author[0000-0002-1691-8217]{Rachael Beaton}
\affiliation{Space Telescope Science Institute, 3700 San Martin Drive, Baltimore, MD 21218, USA}

\author[0000-0001-5201-8374]{Dillon Brout}
\affiliation{Departments of Astronomy and Physics, Boston University, Boston, MA 02215, USA}

\author[0009-0008-4185-8798]{Tianrui Wu}
\affiliation{Department of Physics, Duke University, Durham, NC 27708, USA}

\author[0000-0003-3889-7709]{Graeme E.~Addison}
\affiliation{Department of Physics and Astronomy, Johns Hopkins University, Baltimore, MD 21218, USA}

\author[0000-0001-8839-7206]{Charles Bennett}
\affiliation{Department of Physics and Astronomy, Johns Hopkins University, Baltimore, MD 21218, USA}

\author[0000-0001-8089-4419]{Richard I.~Anderson}
\affiliation{Institute of Physics, \'Ecole Polytechnique F\'ed\'erale de Lausanne (EPFL), Observatoire de Sauverny, 1290 Versoix, Switzerland}

\author[0000-0003-3460-0103]{Alexei V. Filippenko}
\affiliation{Department of Astronomy, University of California, Berkeley, CA 94720-3411, USA}

\author[0000-0003-3460-0103]{Anthony Carr}
\affiliation{School of Mathematics and Physics, University of Queensland, Brisbane, QLD 4072, Australia}
\affiliation{Korea Astronomy and Space Science Institute, Yuseong-gu, Daedeok-daero 776, Daejeon 34055, Republic of Korea}

\begin{abstract}
\vspace{-12pt}
\ \par

We cross-check the {\it HST} Cepheid/SNe~Ia distance ladder, which yields the most precise local H$_0$, against early {\it JWST} subsamples ($\sim$1/4 of the {\it HST} sample) from SH0ES and CCHP, calibrated only with NGC$\,$4258. We find {\it HST} Cepheid distances agree well ($\sim$1$\sigma$) with all combinations of methods, samples, and telescopes. The comparisons explicitly include the  measurement uncertainty of each method in NGC$\,$4258, an oft-neglected but dominant term. Mean differences are $\sim$0.03 mag, far smaller than the 0.18 mag ``Hubble tension.''  Combining all measures produces the strongest constraint yet on the linearity of {\it HST} Cepheid distances, $0.994 \pm 0.010$, ruling out distance-dependent bias or offset as the source of the tension at $\sim$7$\sigma$.

However, current {\it JWST} subsamples produce large sampling differences in H$_0$ whose size and direction we can directly estimate from the full {\it HST} set. We show that $\Delta$H$_0\sim$2.5~\kmsmpc\ between the CCHP {\it JWST} program and the full {\it HST} sample is entirely consistent with differences in sample selection. We combine all {\it JWST} samples into a new distance-limited set of 16 SNe~Ia at $D\leq 25$ Mpc. Using {\it JWST} Cepheids, JAGB, and TRGB, we find  $73.4\pm2.1$, $72.2\pm2.2$, and $72.1\pm2.2$ \kmsmpc, respectively. Explicitly accounting for common SNe, the three-method {\it JWST} result is H$_0=72.6\pm2.0$, similar to H$_0=72.8$ expected from {\it HST} Cepheids in the same galaxies. The small {\it JWST} sample trivially lowers the Hubble tension significance due to small-sample statistics and is not yet competitive with the {\it HST} set (42 SNe~Ia and 4 anchors), which yields 73.2$\pm$0.9. Still, the joint {\it JWST} sample provides important crosschecks which the {\it HST} data passes.
\end{abstract}

\section{Introduction}
\label{sec:intro}

\begin{figure}[t!] 
\begin{center}
\includegraphics[width=0.8\textwidth]{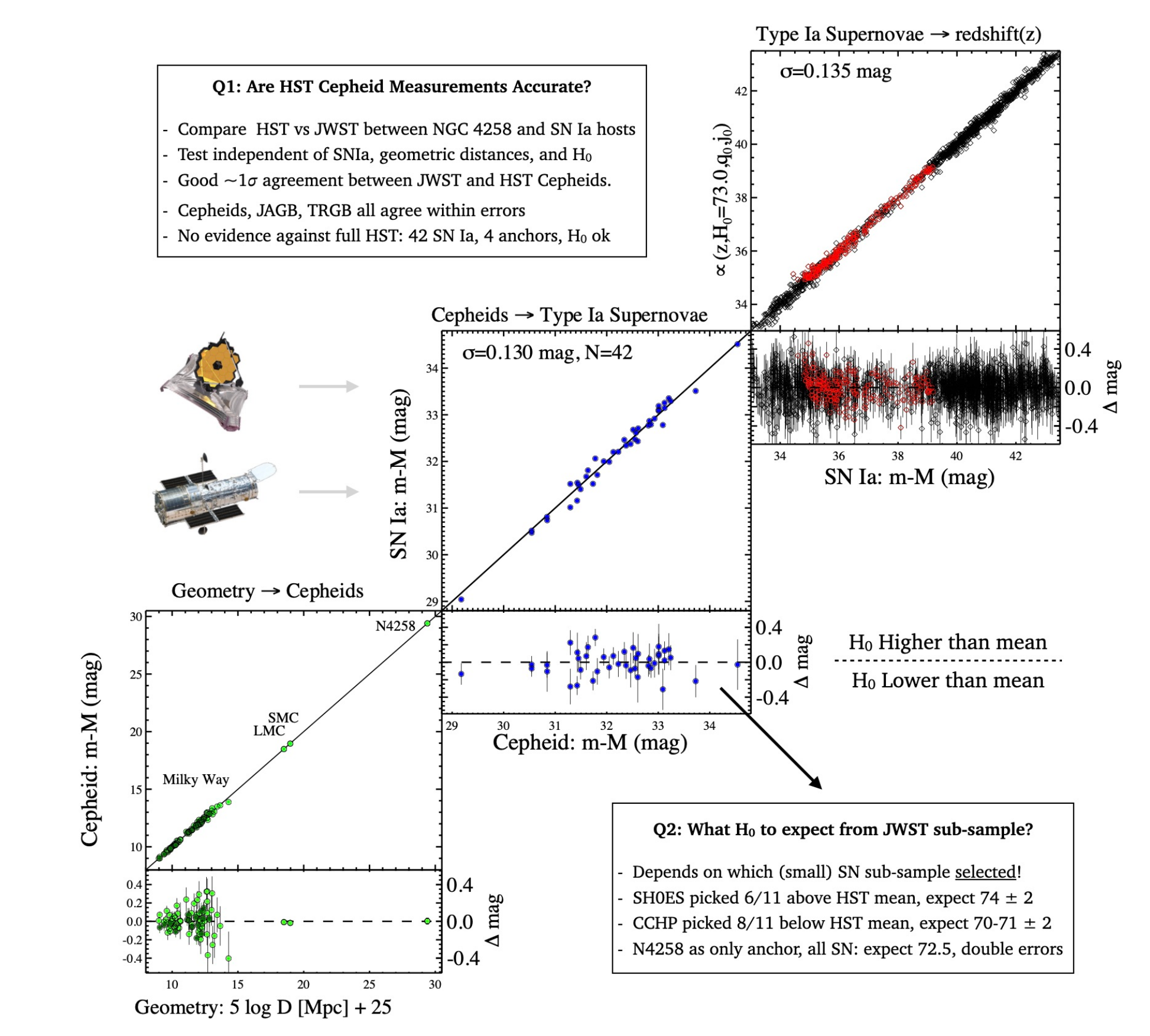}
\end{center}
\caption{A graphic explaining two issues that may be addressed using new {\it JWST} observations, overlaid on top of the three-rung distance ladder presented by \citetalias{Riess:2022}. The first (see \S2) compares relative distances measured between NGC$\,$4258 and SN~Ia hosts by {\it HST} Cepheids with other indicators. This can be done at the 0.03--0.06 mag level depending on the quality of the measurements. This test is independent of SN magnitudes, anchor distances, and the Hubble flow. The second (see \S3) refers to the measurement of H$_0$ that can be expected from the selected SN subsamples and chosen anchor. This is strongly affected by the subsample selection owing to sampling noise and is most meaningfully compared to the same selections made from the full {\it HST} sample as indicated.}
\label{fg:explainer} 
\end{figure}

Currently, the primary route to a $\sim$1\% local determination of the Hubble constant (H$_0$) comes from distance ladders composed of three ``rungs": (1) geometric distance measurements to multiple independent ``anchors"; (2) primary distance indicators (i.e., standard or standardizable luminous stars) observed in these anchors and in the hosts of several dozen nearby Type Ia supernovae (SNe~Ia); and (3) SNe observed in these local hosts and in the Hubble flow. Given state-of-the-art measurements, the precision of this route has been most limited by the sample size, $N$, of the SN~Ia hosts within the range of primary distance indicators such as Cepheid variable stars, tip of the red giant branch (TRGB), Mira variable stars, or C-rich asymptotic giant branch (AGB) stars, scaling as $\sim$(5--8\%)/$\sqrt{N}$ where the numerator depends on the (empirically demonstrated) quality of the SN distances. As of 2022, the largest collection of homogeneously measured SNe~Ia \citep{Brout:2022, Scolnic:2022} is complete to $D \leq 40$ Mpc or redshift $z \leq 0.01$. It consists of 42 SNe~Ia in 37 host galaxies calibrated with observations of Cepheids with the {\it Hubble Space Telescope (HST)}, the heritage of more than 1000 orbits (a comparable number of hours) invested over the last $\sim$20~yr \citep[][hereafter \citetalias{Riess:2022}; see Fig.~\ref{fg:explainer}]{Riess:2022}. The size of this sample reduces fluctuations in H$_0$ due to the intrinsic dispersion of SN magnitudes to $\leq 1$\%. The sample of four anchors --- NGC$\,$4258, the Milky Way (MW), and the Large and Small Magellanic Clouds (LMC and SMC) --- all observed with {\it HST}, also reduces geometric calibration errors to $\leq 1$\% \citep{Breuval:2024}.

However, the ``Hubble tension,'' a decade-long discrepancy now reaching a $>$5$\sigma$ difference between the local determination of H$_0$ and the prediction from $\Lambda$CDM calibrated by the cosmic microwave background (CMB) and which may augur new physics, has motivated additional tests and cross-checks \citep[for an extensive and recent review of tests, checks and independent measures, see][]{DiValentino:2021, Tully:2023, Verde:2023}. The new capabilities of the {\it James Webb Space Telescope (JWST)} offer the means for additional cross-checks by comparing distances measured to SN host galaxies with those measured with {\it HST}. Two {\it JWST} programs received time in Cycle 1 to (re-)measure distances between one anchor, NGC$\,$4258, and a subsample of 5--10 hosts of 8-11 SNe~Ia using a variety of established and new primary distance indicators: Cepheids, TRGB, and C-rich AGB stars. In this analysis, we review the consistency between all of the {\it JWST} and {\it HST}-measured host distances. From this comparison we derived the strongest constraints on the linearity of measured {\it HST} Cepheid distances. We also analyze an independent and important issue: the fluctuations in $H_0$ due to the small size of the SN samples calibrated with {\it JWST} if {\it exclusively} used to determine the parameter. 

{\it JWST} has certain distinct advantages (and some disadvantages) compared to {\it HST} for measuring distances to nearby galaxies. For Cepheids, {\it JWST} offers a $\sim$2.5$\times$~higher near-infrared (NIR) resolution than {\it HST} to mitigate crowding, though an optical facility like {\it HST} is still needed to find Cepheids. The observed reduction in the scatter of the Cepheid Period-Luminosity relation ($P$--$L$) for ideal observations with {\it JWST} is remarkable \citep{Riess:2024}. For the TRGB distance indicator, {\it JWST} offers greater depth in the NIR where red giants are relatively brighter than other stars, but at the cost of a large and uncertain color dependency for the method. An $I$-like band is usually the choice for TRGB measurements because this feature appears flat in a color-magnitude diagram (CMD) of stars, where theory indicates differences in metallicity and age move stars primarily in color rather than brightness (at least with low to intermediate metallicities; \citealt{2020ApJ...891...57F}). In the NIR, theory shows that the TRGB is tilted and non-linear, and its shape is more sensitive to age and metallicity, making such measurements more challenging \citep{2014AJ....148....7W,McQuinn:2019,Madore:2023,Hoyt:2024a,Newman:2024}. Another promising, newly employed ``standard population," C-rich AGB stars (JAGB), produce a clump of stars in the NIR CMD with a luminosity function that may be measured and is presumed to be standard. The techniques for the newer methods are still evolving \citep[][and more]{Dalcanton:2012, Beaton:2018, Durbin:2020, Madore:2020, Ripoche:2020, Parada:2021, Lee:2023, Hoyt:2024a, Li:2024}. 

Even by observing only small subsamples of SN~Ia hosts and NGC$\,$4258, {\it JWST} is able to provide a strong crosscheck of distances in the first two rungs because this comparison is independent of SNe~Ia and their own intrinsic scatter, empirically $\sim$0.12--0.19~mag. This test is also independent of the knowledge of the geometric distance to the anchor. It is, however, still subject to the uncertainties in the measurements of both distance indicators or both telescopes in NGC$\,$4258, an error floor to the comparison of two indicators or telescopes and which is not reduced by adding more SN hosts. We explain this cross-check in Fig.~\ref{fg:explainer}. 

Several comparisons of {\it the measured difference in distance} between NGC$\,$4258 and a subsample of SN~Ia hosts between {\it HST} Cepheids and {\it JWST}-measured have already been presented. \cite{Riess:2024} found agreement between $>$1000 Cepheids measured with {\it JWST} and {\it HST} for 5 SN~Ia hosts from the SH0ES team, when both sets were analyzed {\it at the same wavelength and with the same slope} for the \PL and using multiple {\it JWST} epochs to measure light-curve phases, with a difference of $-0.011\pm0.032$~mag. \cite{Li:2024} compared the $I$-band ($F090W$) TRGB between NGC$\,$4258 measured by \cite{Anand:2024} and 8 hosts of 10 SN~Ia in the SH0ES {\it JWST} sample, finding agreement with {\it HST} Cepheids at the level of $0.01 \pm 0.04$ (stat) $\pm 0.04$ (sys)~mag. A direct comparison of the 11 $I$-band TRGB distances to SN~Ia hosts from {\it HST} by \cite{Freedman:2021} and by R22, relative to NGC$\,$4258 (where they define $M_I=-4.05$ mag; \citealt{Jang:2021}), produces a difference of $0.01\pm0.04$~mag\footnote{A review by \citet{Freedman:2023} notes the ``excellent agreement between the published Cepheid distances in \cite{Riess:2022}\\and TRGB distances in \cite{Freedman:2019}, which in the mean, agree to 0.007 mag."}. \cite{Li:2024} compared JAGB measurements with {\it JWST} between 4 SN~Ia hosts and NGC$\,$4258, finding agreement with Cepheids, though the precision was limited by systematic differences in characterizing the JAGB luminosity function (LF). A study by \citet[][hereafter F24]{Freedman:2024} and \citet[][hereafter L24]{Lee:2024} compares 3 distance indicators measured with {\it JWST}, but to each other rather than to the {\it HST} R22 Cepheid sample, the results of which we will address in the next section. We review the comparisons of each method, telescope, and group, using the {\it HST} Cepheid R22 sample as a reference since it contains all of the other measured subsamples.
While fluctuations may occur for any selected subsample, it is important and the convention to make use of a large, distance-limited sample to reduce bias and fluctuations. It is important to note that neither the {\it JWST} SH0ES nor the CCHP targets constitute a distance-limited sample to the distance of their farthest target. Remarkably (given the lack of program coordination), a distance limited sample is formed by the {\it combined} {\it JWST} observations of the two programs for $D \leq 25$ Mpc, which includes 16 SNe~Ia in that range.

It seems reasonable to use {\it JWST} observations to compare relative distance methods and sources ({\it HST} Cepheids, {\it JWST} Cepheids, I-TRGB, NIR-TRGB, JAGB) to identify which are in mean accord ($< 2$--2.5$\sigma$ is conventional) and which are not ($>2.5$--3.0$\sigma$ is problematic). If {\it HST} Cepheid measurements are in accord with others, there would be no evidence to reject their use, \textit{or} the large sample of SNe~Ia calibrated with {\it HST} to determine H$_0$ and which offer the only route to a local measurement approaching 1\% precision. 

It is less clear that there is value (beyond the host distance cross-checks) in measuring H$_0$ exclusively from these {\it JWST}-selected subsamples, because their small size quadruples the variance in H$_0$ owing to the intrinsic dispersion of SNe~Ia while introducing no new objects. This SN sampling noise is already reduced within the {\it HST} full sample of 42 SN Ia and would otherwise revert to past levels \citep[at the level of][]{Riess:2011} with {\it JWST} if it is the only source used to determine H$_0$. Furthermore, the selected subsamples cannot be considered without the potential for observer bias \citep[e.g., the SN absolute magnitudes having been previously measured, most by][]{Riess:2016,Riess:2022}, nor are they complete in distance. 

One further consideration is the availability of geometric anchors used to determine H$_0$. There are four geometric anchors \citep{Riess:2022, Breuval:2024} for Cepheids directly observed by {\it HST}. For population-based methods (e.g., TRGB or JAGB), parallax is less useful and 3 of the 4 anchors are too bright for {\it JWST}. It is not yet clear if {\it JWST} will be able to usefully observe indicators in the MW, LMC, or SMC, a feat {\it HST} achieved only through rapid spatial scanning and gyro-guided fast slews \citep{Riess:2018,Riess:2019}, not currently available for {\it JWST}. From {\it HST} Cepheids (\citetalias{Riess:2022}), there are predictable fluctuations in H$_0$ based solely on anchor choice. NGC$\,$4258, the sole anchor available for {\it JWST} produces the lowest value of H$_0$ of the 4 anchors, decreasing H$_0$ from the 4-anchor mean of \citet{Breuval:2024} by $\Delta H_0 \approx 0.7$~\kmsmpc. Reducing anchors also increases the uncertainty. With NGC$\,$4258 as the sole anchor, R22 found H$_0=72.5 \pm 1.5$~\kmsmpc\ (a reduction to a 3$\sigma$ tension with CMB+$\Lambda$CDM even before any reduction in SN sample size). For a sample of only $\sim$10 SNe with {\it JWST}, combining the error in {\it measuring primary distance indicators in} NGC$\,$4258 ($\sigma \approx 1.3$~\kmsmpc) with the SN subsample ($\sigma \approx 1.5$~\kmsmpc) yields $\sigma \approx 2.0$~\kmsmpc\ before including the geometric calibration ($\sigma \approx 1.1$~\kmsmpc), thus a total error of 2.3~\kmsmpc\ which would be insufficient to detect the present tension of $\Delta H_0 \approx 5$--6~\kmsmpc. A small, $D\leq 25$ Mpc distance-limited sample from R22 with only NGC$\,$4258 as anchor selected from the {\it HST} R22 samples gives H$_0=72.3 \pm 1.8$ \kmsmpc, a reduction to a marginal 2.5$\sigma$ tension (i.e., assuming perfect measurement agreement with {\it JWST}). Simply put, slashing the size of the distance-ladder sample trivially reduces the significance of the tension by inflating uncertainties rather than explaining it. A valid assessment of the tension would require taking into account all relevant data while correcting subsample bias.

Here we analyze early {\it JWST} observing programs and the distance measurements from each. The consistency of the various distance probes are in \S 2, of  H$_0$ in \S3, and we discuss notable differences with the analysis of some of this data in F24 in Appendix B. 

\begin{figure}[t!] 
\begin{center}
\includegraphics[width=0.65\textwidth]{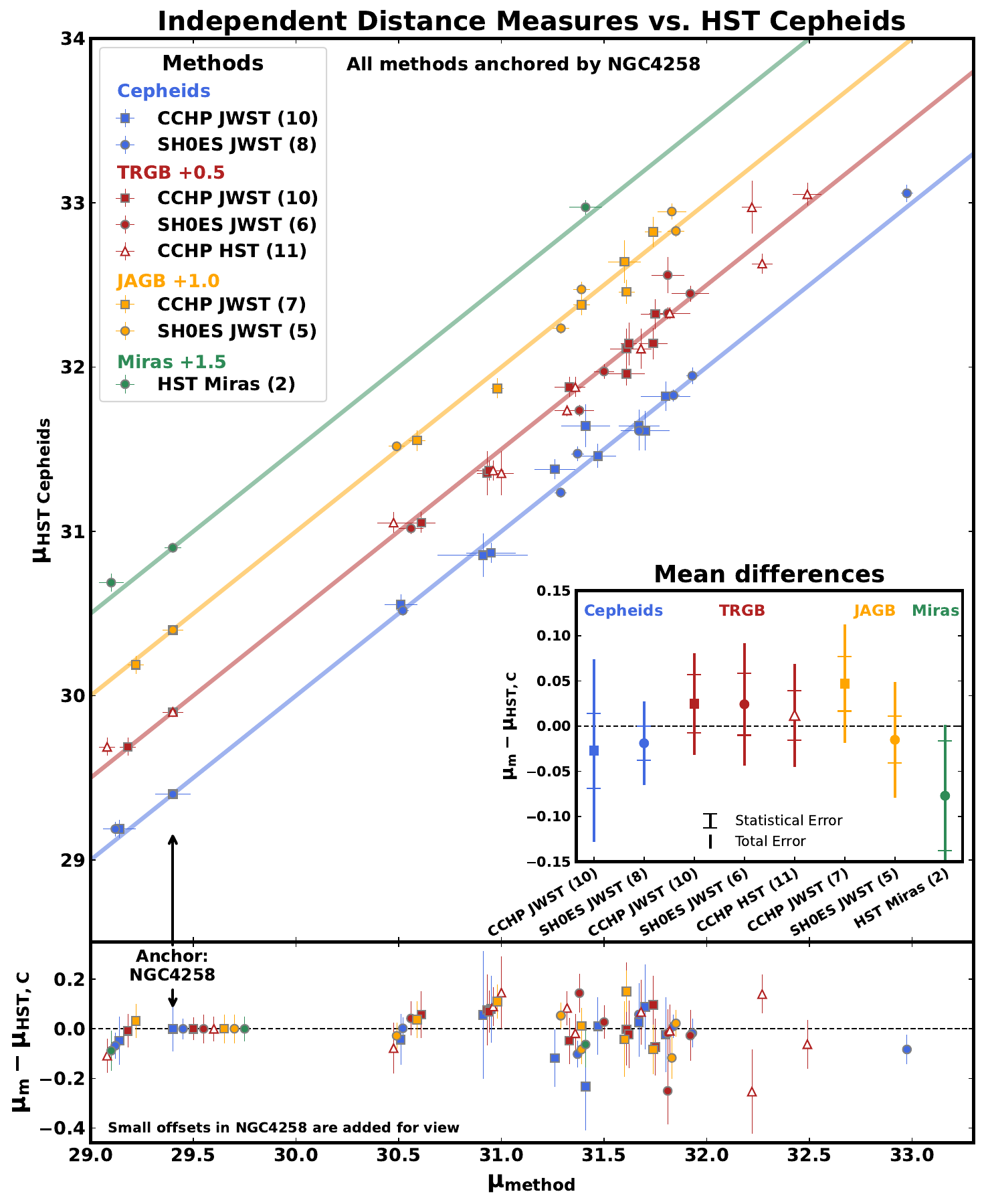}
\end{center}
\caption{The main panel compares SN Ia host distances calibrated with NGC 4258 for various distance methods, samples, and telescopes (X axis) and HST Cepheids (Y axis). 
The lower panel shows the differences between these on a per host basis. The inset shows the  mean differences for the whole samples.
{\it HST} Cepheids are all observed by {\it HST} and analyzed by \citetalias{Riess:2022} and \cite{Riess:2024}; see Table \ref{tab:distpar} with the mean results given in Table~\ref{table:distance_comparisons}. {\it JWST} (and specific {\it HST}) results can be found either in R24 (corresponding to SH0ES-selected) or F24 (corresponding to CCHP-selected) and computed here from the distances table, Table \ref{tab:matrix}. All measures are in good, $\sim$1$\sigma$ agreement. The largest uncertainties in these comparisons arise from the individual measurements in NGC$\,$4258 and the mean measures of the SN~Ia hosts as given in Table \ref{tab:comperr}.}
\label{fg:comparedis} 
\end{figure}

\section{Comparing Distance Measurements}
\subsection{Data}

\subsubsection{Previous {\it HST} Samples}

The full {\it HST} Cepheid sample from SH0ES (R22) is a distance-limited ($z \leq 0.01$) set of 42 SNe~Ia in 37 late-type hosts (since 2021) calibrated by 4 different anchors. This sample provides an ideal reference for all other samples, because it contains all of the subsamples described here. In Table~\ref{tab:distpar} we present a compilation of relevant information for the full {\it HST} sample of SN~Ia hosts, and selected {\it JWST} subsamples, similar to Table 6 of R22 but with NGC$\,$4258 as the sole anchor\footnote{The mean change in distance modulus when using only NGC$\,$4258 instead of the 3 anchors in R22 is an increase of 0.012 mag, but the actual value varies by host (based on fit 10 of R22). As discussed by R22, these measures are approximations to the full, simultaneous ladder fit, as they do not use the full covariance approach (but they are consistent with the full fit at the $\sim$0.01~mag level).}, allowing for a direct comparison with other methods or telescopes that rely only on this galaxy. We also make use of {\it HST} $I$-band TRGB measurements in SN~Ia hosts given by \cite{Freedman:2021} calibrated to NGC$\,$4258 and Miras measured in SN~Ia hosts calibrated to NGC$\,$4258 \citep{Huang:2024}.

\subsubsection{JWST Observations}

Several early {\it JWST} programs reobserved a number of hosts from the the full {\it HST} sample. {\it JWST} Cycle 1 GO–1685 (\citealt{2021jwst.prop.1685R}; SH0ES) observed 5 hosts of 8 SNe~Ia in two epochs with NIRCAM filters $F090W$, $F150W$, and $F277W$ to measure Cepheids and the I-TRGB (later adding the C-rich AGB star candle). Their choice of hosts favored those with the largest numbers of Cepheids and SNe~Ia per host. 
GO–1995 (\citealt{2021jwst.prop.1995F}; CCHP) observed 10 hosts of 11 SNe~Ia in one epoch and NIRCAM filters $F115W$ and $F444W$ (part way changing $F444W$ to $F356W$) to measure Cepheids, the NIR-TRGB, and carbon stars. Their choice of hosts favored suitability for the three chosen distance indicators in all hosts, resulting in targets at $D\leq 25$~Mpc. Their most distant galaxy, NGC$\,$4639 ($\mu=31.823 \pm 0.091$ in R22 or Table 7), is consistent with the high end of this range. One host (of two SNe~Ia), NGC$\,$5643, was selected by both programs. Cycle 1 program GO-2581 \citep{2021jwst.prop.2581C} serendipitously observed one R22 host, NGC$\,$4038, where we recovered 22 Cepheids in the same filter $F150W$ used to measure Cepheid distances in the SH0ES baseline. An early Cycle 2 program GO–4087 \citep{2023jwst.prop.4087H} observed 90 Cepheids in the SH0ES host M101 with the same $F150W$ filter. We will consider these two hosts (NGC$\,$4038 and M101, both also present in the CCHP selected sample) here as part of a {\it JWST} SH0ES selected/system sample because they can be measured with the exact same NIRCAM $F150W$ system as the others. A {\it JWST} Cycle 2 program, GO-2875 (\citealt{2023jwst.prop.2875R}; SH0ES) has observed another 5 hosts of 7 SNe~Ia. One of these, NGC$\,$3447, is at $D\leq25$~Mpc based on {\it HST} Cepheid measurements (R22), and would complete the distance-limited SN Ia sample with {\it JWST} Cepheid observations. We measured 140 Cepheids in this host (following the same procedures given in R24) and include its measurement in Table~\ref{tab:matrix}. We also include {\it JWST} TRGB measurements presented in \cite{Li:2024b} with $D \leq 25$ Mpc. All distance data are in Table \ref{tab:matrix}. We note that both groups use the same primary software for measuring stellar photometry (DOLPHOT; \citealt{2016ascl.soft08013D, 2024ApJS..271...47W}) so there is every reason to expect measurement consistency. Here we adopt the distance measurements provided in F24 at face-value.

There are several differences in how the observations were obtained and in the analysis procedures adopted. Describing these differences in detail goes beyond the scope of this paper; for full details, we direct the reader to the original papers \citep{Riess:2023, Riess:2024, Anand:2024, Li:2024, Li:2024b, Freedman:2024, Lee:2024}. One specific difference impacting the quality of the results is the choice of primary filter and the number of epochs; GO-1685 adopted a redder filter, $F150W$, and obtained observations at two different epochs to constrain the phases of individual Cepheids \citep{Riess:2024}, while GO-1995 used $F115W$ as primary and obtained a single epoch \citep{Freedman:2024}. For the one host in common, NGC$\,$5643, the $P$--$L$ relation obtained in $F150W$ using the two-epoch phase constraints has a scatter of 0.17~mag, while the relation obtained using a single epoch of $F115W$ has a scatter of 0.23~mag, about 35\% larger. Both yield similar distances (SH0ES, $\mu=30.52$~mag; CCHP, $\mu=30.51$~mag).  \footnote{We also note that the distance uncertainties provided in F24 for their Cepheid measurements are a factor of a few times larger than would be expected by conventional error propagation, i.e., using the error in the mean of the \PL relation (based on the empirical dispersion) or what was estimated for the same galaxies and Cepheids measured in the HST Key Project \cite{Freedman:2001} or in R22.  Rather, the addition of {\it JWST} data should reduce the distance uncertainty.  For example for NGC 4536 F24 gives $\sigma=0.12$ mag while the HST KP gives $\sigma=0.04$ mag and R22 found $\sigma=0.05$.  The cause appears to be that F24 states that they employ the dispersion of the \PL as the uncertainty in the \PL intercept rather than the error in the mean which will be far greater.  We do not know why and this practice appears inconsistent with F24's use of the error in the mean for JAGB and similar for TRGB.  We caution the F24 Cepheid error estimates are almost certainty overestimated and may undervalue a $\chi^2$ comparison to other distance indicators.}

\subsection{Comparison Formalism}

To briefly review, we can measure the distance modulus $\mu^0_i$ to an SN~Ia host by measuring the magnitude ($m$) or dereddened magnitude ($m^0$) of a type of standard candle, $a$ or $b$, in the $i$th host and also in NGC$\,$4258. From these, we obtain their difference in distance modulus and then add the known, geometric (maser) distance modulus, $\mu_{0,{\rm N}4258}$:

\begin{equation}
    \mu^0_{i,a}=m^0_{i,a}-m^0_{{\rm N}4258,a}+\mu_{0,{\rm N}4258}+{\rm sys}_a\, ,
\end{equation}
\begin{equation}
    \mu^0_{i,b}=m^0_{i,b}-m^0_{{\rm N}4258,b}+\mu_{0,{\rm N}4258}+{\rm sys}_b\, ,
\end{equation}
\begin{equation}
    \Delta\mu({a-b})=\mu^0_{i,a}-\mu^0_{i,b}=(m^0_{i,a}-m^0_{i,b})-(m^0_{{\rm N}4258,a}-m^0_{{\rm N}4258,b})+({\rm sys}_a-{\rm sys}_b)\, ,
\end{equation}
with
\begin{equation}
    \sigma(\Delta\mu({a-b}))^2=\sigma_a^2+\sigma_b^2+\sigma(m^0_{{\rm N}4258,a})^2+\sigma(m^0_{{\rm N}4258,b})^2+\sigma({\rm sys}_a)^2+\sigma({\rm sys}_b)^2\, ,
\end{equation}
where
\begin{equation}
\sigma_a^2=1/\sum^n_{i=1}\sigma(m_{i,a})^{-2} \ \textrm{and}\ \sigma_b^2=1/\sum^n_{i=1}\sigma(m_{i,b})^{-2}.\label{eq:errors}
\end{equation}
Note that these equations assume that \textit{all} error terms are independent and uncorrelated.  We note that an individual host magnitude uncertainty, $\sigma(m_{i,a})$ is defined to include both measurement error and intrinsic scatter of the candle.  Equation 4 provides the error in the difference in distance moduli for a single host or for a sample of hosts by substituting $\sigma_a$ and $\sigma_b$ from equation 5.

The difference in distance modulus between the methods, $\Delta\mu(a-b)$, is independent of knowledge of the geometric distance, $\mu_{0,{\rm N}4258}$.
For a sample of $N$ SN~Ia hosts, the uncertainty in the measured differences, $(m^0_{i,a}-m^0_{i,b})$, 
will be reduced by $\sqrt{N}$ (Eq.~\ref{eq:errors}), but uncertainties from the independent measurements of two methods in NGC$\,$4258, $(m^0_{{\rm N}4258,a}-m^0_{{\rm N}4258,b})$, become the limiting factors for measuring the difference. We see no way to avoid this term as it encompasses the quality of the measurements in NGC$\,$4258 which could vary from excellent to poor which would be reflected in this term. For a population-based indicator like TRGB or JAGB, this term would also include any intrinsic field-to-field (or population to population) scatter. F24 (Table 5) refers to measurement uncertainties in NGC$\,$4258 together with the 1.5\% geometric distance uncertainty from \cite{Reid:2019} as a systematic uncertainty in $H_0$, terms which total 2.2\% to 4.3\% for each method.

An (external) systematic error for each method, sys$_a$ and sys$_b$, may describe a relative uncertainty between NGC$\,$4258 and the SN hosts and may (be hoped to) partially cancel if it applies equally to method $a$ and $b$ (or may have a covariance term to include).
The uncertainty in the two-method comparison thus contains 3 basic terms: the error in the host subsample mean measurements, the uncertainties in each method's measurements made in NGC$\,$4258, and any relevant (external) systematic error that applies between them. These terms apply separately to both telescopes or methods and thus affect their comparison. 

We give illustrative but realistic error budgets for the methods and comparisons in Table \ref{tab:comperr}. For each method, we derive the statistical uncertainties from the measurements in the first and second rungs for an assumption of one anchor (NGC$\,$4258) and $\sim$10 SN hosts. We also give a systematic uncertainty due to differences between the SN host and NGC$\,$4258 applicable for that method. 

For a realistic measurement error per host for Cepheids, TRGB, or JAGB of 0.05--0.08 mag, 10 hosts will yield an error in the mean (equation 5) of 0.02-0.03 mag. A critical and often-neglected term in a two-method comparison is the measurement error for NGC$\,$4258 (Cepheid $P$--$L$ intercept or TRGB/JAGB apparent magnitude) for each telescope and method, which does {\it not} cancel in the comparison in the way the geometric distance error does. Ideally NGC$\,$4258 would be the best-measured object, but in practice this may not occur. For the  quality of Cepheid measurements from {\it JWST} presented in R24, the uncertainty in the intercept of NGC$\,$4258 using $\sim$100 Cepheids is 0.02 mag though appears greater for the CCHP observations (F24, Table 5). For TRGB, we take the measurement error from F24 of 0.035 mag (see Appendix B for further discussion). For JAGB, we take this uncertainty to be 0.05 mag from F24 (Table 5) and reported by \citet{Lee:2024}.

The primary (external) systematic difference in the case of Cepheids derives from the difference in period distribution, together with the uncertainty in the slope of the assumed $P$--$L$ relation\footnote{\cite{Riess:2024} previously presented a mean {\it JWST}--{\it HST} difference for 5 hosts and $\sim$1000 Cepheids. The systematic error term from the $P$--$L$ slope uncertainty was reduced in this comparison, $({\rm sys}_a-{\rm sys}_b)$, by measuring both the {\it HST} and {\it JWST} samples at the same wavelength (hence the same error for $a$ and $b$) which also allows for fixing both to the same $P$--$L$ slope.}. Metallicity effects largely cancel out, as NGC$\,$4258 and the SN~Ia host subsample have very similar measured metallicities; we adopt a common metallicity term of 0.2 mag\,dex$^{-1}$, accurate to $\sim$0.02 mag after application. For a subsample of $\sim$10 hosts, random uncertainties will depend on the specific targets selected, including the number of Cepheids in each target, and the number of epochs observed, which in turn impacts the $P$--$L$ scatter; individual uncertainties vary by a factor of a few from target to target. The systematic difference for TRGB is due to intrinsic TRGB variation between the SN hosts and NGC$\,$4258. This is discussed at length by \cite{Wu2023}, \cite{Scolnic:2022} and \cite{Anderson:2024}, and will be different for TRGB measured in the optical or NIR. We give a range of 0.01--0.08 mag. For JAGB, the range of 0.01--0.10 mag is given owing to the nature of choices in the analysis. As discussed by \cite{Li:2024}, there are variations up to the $0.1$ mag level when applying different statistical techniques (e.g., mode/median).

 We find the statistical error from a comparison is likely to be $\sim$0.06 mag but can range from 0.03-0.10 mag for the {\it JWST} sample, and the systematic contribution is 0.03--0.1 mag. For each technique, the uncertainty of the measurement for each method in NGC$\,$4258 is the dominant source.

 F24 gives uncertainties in two-method comparisons which are a factor of $\sim3$ smaller than those we derive from their data following standard error propagation (equation 4 and Table 1). We discuss this further in the Appendix B where we trace the origin to the non-propagation of the given measurement uncertainties (or any uncertainty) in the measurements of each distance indicator in NGC$\,$4258.

\subsection{Distance Comparisons} 

We show in Fig.~\ref{fg:comparedis} and Table~\ref{table:distance_comparisons} a comparison of distance measurements using {\it HST} Cepheid observations from \citetalias{Riess:2022} and those from other {\it HST} and {\it JWST} observations/techniques. We use the uncertainties as derived in Table~\ref{tab:comperr} for the comparisons. 
We find that all techniques are in good accord at the $\sim$1$\sigma$ and an average level of $\sim$0.03 mag. 

When comparing Cepheid measurements from {\it JWST} and {\it HST}, we see that the mean difference with the CCHP {\it JWST} Cepheid sample of 10 hosts  is $-0.027 \pm 0.10$ mag, in the sense of {\it JWST} being closer. We note that the CCHP measurements are obtained in $F115W$, a different filter than used with {\it HST}, so systematic error cancellation does not apply which is part of the reason for the relatively large error. See Fig.~1 of \cite{Riess:2023} for sources of $P$--$L$ dispersion.
When we compare to the Cepheid measurements of the 8 SH0ES-selected hosts of 11 SNe~Ia, we find a mean difference with the {\it HST} Cepheids of $-0.02 \pm 0.03$ mag. The uncertainty here is smaller owing to Cepheid samples in the SH0ES hosts which are a factor of $\sim$3 larger, \PL scatter which is 40\% smaller due to multiple epochs and colors, and the matching of wavelength between {\it JWST} and {\it HST} (and hence the \PL slope).  The {\it JWST} and {\it HST} Cepheid comparisons presented by \cite{Riess:2024} included 15 variants (i.e., choices) including no period limits, $P>15$~d limits, steeper and shallower $P$--$L$ slope, no or double metallicity correction, most crowded and least crowded halves, no phase correction or single random phase, and the use of Cepheid colors (for dereddening) from {\it HST} or from {\it JWST} NIR or {\it JWST} mid-IR. The mean differences ranged from $-0.030$ to 0.017~mag, with the fluctuations all less than 1$\sigma$. We conclude that {\it HST} and {\it JWST} Cepheid measurements are robustly consistent (which is also true of the individual team samples). 

For comparisons between {\it JWST} JAGB and {\it HST} Cepheids, we measure a difference with the CCHP {\it JWST} sample of $0.047 \pm 0.066$ mag, and for the SH0ES {\it JWST} sample we measure a difference of $-0.015 \pm 0.064$ mag, though this can vary between 0.02 and $-0.04$ depending on whether the JAGB mode, median or mean statistic is used which remains an arbitrary choice \citep{Li:2024}. L24 find that if they apply the mean or median instead of the mode, the JAGB distances would decrease (due to the greater skew of NGC$\,$4258); the agreement would then be $0.01 \pm 0.055$~mag. F24 noted a larger difference of 0.08~mag between Cepheids and JAGB, but 
the Cepheid reference was their {\it JWST} Cepheids, not the {\it HST} Cepheids compared here. As discussed in Appendix B, with the measurement errors in NGC$\,$4258 included, even the {\it JWST}-to-{\it JWST} difference is only 1$\sigma$. The CCHP {\it JWST} JAGB-Cepheid difference also matches the size and direction of the difference between CCHP JAGB and CCHP {\it HST} TRGB in common, 0.07 mag. 

When we compare {\it HST} Cepheid distances to TRGB, we measure differences of 0.032, 0.02, and 0.01~mag for {\it JWST} CCHP, {\it JWST} SH0ES, and {\it HST} CCHP, respectively, each with a total uncertainty of $\sim$0.05$-$0.06 mag.
 The {\it JWST} I-TRGB measurements are presented by \cite{Li:2024b}, and here limited to the 6 hosts of 8 SNe~Ia with $D\leq 25$ Mpc. All of these comparisons are in good accord at the $\sim$1$\sigma$ level.
\begin{deluxetable*}{lrr}[t!]
\tabletypesize{\small}
\tablewidth{0pt}
\tablecaption{Representative Error Budget for Comparing two {\it JWST} Distance Method Samples ($\sim$10 hosts calibrated to NGC$\,$4258) \label{tab:comperr}}
\tablehead{
\multicolumn{1}{l}{Term} & \colhead{$\sigma(\textrm{stat})$} & \colhead{\hspace*{0.75in}$\sigma(\textrm{sys})$}\\[-3pt]
& \multicolumn{2}{c}{[mag]}}
\startdata
\hline
\textbf{Cepheids} & & \\
\hline
Method Calibration Measurement: intercept in NGC$\,$4258 ($\sim$100 Cepheids) & 0.03 & \multicolumn{1}{c}{--}  \\
SN Host Sample means: intercepts, $[\sim \sigma(P-L) / \langle N_{\textrm{Ceph}}\rangle^{1/2}] \times N_{\textrm{ hosts}}^{-1/2}$\hspace*{0.75in} & 0.02 & \multicolumn{1}{c}{--} \\
Systematic between SN hosts and NGC$\,$4258, $P$--$L$ slope: $\Delta \log\,P\ \times\ \sim 0.3 \sigma$(slope), other (a) & \multicolumn{1}{c}{--} & 0.01-0.05 \\
\hline 
Cepheid Subtotal & 0.04 & 0.03 \\
~ & ~ & ~ \\
\hline
\textbf{TRGB} & & \\
\hline
Method Calibration Measurement: TRGB mag in NGC$\,$4258  & 0.04 & \multicolumn{1}{c}{--} \\
SN Host Sample means: $\sim$0.06$\times N_\textrm{hosts}^{-1/2}$ & 0.02 & \multicolumn{1}{c}{--} \\
Systematic between SN hosts and NGC$\,$4258, Intrinsic TRGB variation (b)& \multicolumn{1}{c}{--} & 0.01-0.08 \\
\hline
TRGB Subtotal & 0.045 & 0.01-0.08 \\
~ & ~ & ~ \\
\hline
\textbf{JAGB} & & \\
\hline
Method Calibration Measurement: JAGB mag in NGC$\,$4258  & 0.05 & 
\multicolumn{1}{c}{--} \\
SN Host Sample means: $\sim$0.06$\times N_\textrm{hosts}^{-1/2}$ & 0.02 & \multicolumn{1}{c}{--} \\
Systematic between SN hosts and NGC$\,$4258, Intrinsic JAGB Method (c) & \multicolumn{1}{c}{--} & 0.01-0.10\\
\hline
JAGB Subtotal & 0.055 & 0.01-0.10 \\
~ & ~ & ~ \\
\hline
\textbf{Comparison between two methods} (d) & 0.06 & $\sim$0.03-0.10 \\
\textbf{Total} & 0.07-0.12 &  \\
\enddata
\tablecomments{An uncertainty budget for the present analysis. Uncertainties are listed separately for each distance method (Cepheids, TRGB, JAGB) and their individual subtotals are given. (a) a mean difference in $\log\,P$ between NGC$\,$4258 and an SN~Ia host of $\sim$0.3 times a slope uncertainty of $\sim$0.05; ``other'' leaves room for other choices but R24 show a range of $<$0.03 mag for a wide range of choices. (b) Term highly dependent on the bands and colors used to measure; the low end would be suitable for $I$-band where TRGB is least sensitive to metallicity and age and the high end for NIR.(c) Level of differences for different measures, e.g., mode vs. mean vs. median of JAGB LF.(d) Depends on (a) and (b), which also will impact the weights; optimistic case here.}
\end{deluxetable*}
\begin{deluxetable}{llrr}
\tablecaption{Comparing SN~Ia Host Distances \label{table:distance_comparisons}}
\tablehead{
\multicolumn{1}{l}{Sample\hspace*{2in}} & \colhead{Team\hspace*{0.5in}} & \multicolumn{1}{c}{$\mu_{\it JWST}$} & \multicolumn{1}{c}{$\sigma$}\\[-6pt]
& & \multicolumn{1}{c}{$ - \mu_{\it HST}$} & \\[-2pt]
& & \multicolumn{2}{c}{[mag]}}
\startdata
\hline
\multicolumn{4}{c}{{\it HST} Cepheids vs. {\it JWST} Cepheids}\\
\hline
10 Hosts       & CCHP  & $-0.027$& 0.10 \\
8 Hosts ($D\leq 25$ Mpc) & SH0ES & $-0.02$ & 0.03 \\
14 Unique Hosts       & Both & $-0.02$ & 0.03 \\
\hline
\multicolumn{4}{c}{{\it HST} Cepheids vs. {\it JWST} JAGB} \\
\hline
7 Hosts (mode)              & CCHP  & 0.047  &  0.066 \\
5 Hosts              & SH0ES & -0.015   &  0.064 \\
\hline
\multicolumn{4}{c}{{\it HST} Cepheids vs. {\it JWST} NIR-TRGB} \\
\hline
10 Hosts             & CCHP  &0.032 & 0.056 \\
\hline
\multicolumn{4}{c}{{\it HST} Cepheids vs. {\it JWST} I-TRGB} \\
\hline
6 Hosts ($D\leq 25$ Mpc)  & SH0ES  & 0.02 & 0.067 \\
\hline
\multicolumn{4}{c}{{\it HST} Cepheids vs. I-TRGB} \\
\hline
11 Hosts              & CCHP & $0.01$ & 0.057 \\
\hline
\multicolumn{4}{c}{{\it HST} Cepheids vs. {\it HST} Miras} \\
\hline
2 Hosts                &    & $-0.08$ & 0.078 \\
\hline
\enddata
\tablecomments{(*): SH0ES $D\leq 25$ Mpc sample includes NGC$\,$4038, M101 and NGC$\,$3447 and excludes NGC$\,$5468 as discussed in text. CCHP distances from F24 Table 2 and including uncertainties in F24 Table 5. Errors in last column follow from equation 4 and use the minimum systematic uncertainties listed in Table 1.}
\end{deluxetable}
\subsubsection{Cepheid Distance Linearity}
\begin{figure}[t!] 
\begin{center}
\includegraphics[width=0.65\textwidth]{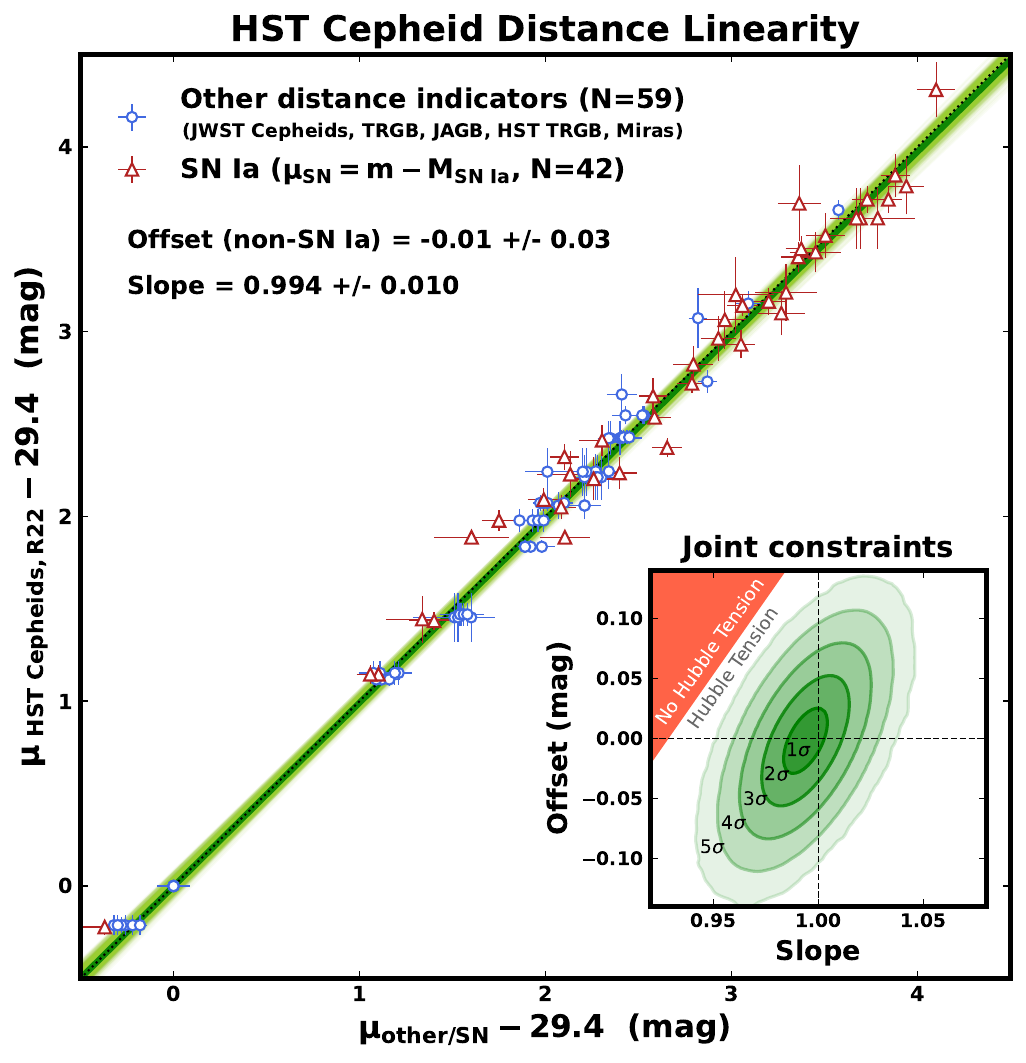}
\end{center}
\caption{{\it HST} Cepheid distance linearity assessed by comparing to all other indicators. The comparison is a linear fit $\mu_{\it HST}={\rm slope}\times \mu_{\it other}-{\it offset}$ between each method (non-SN) and {\it HST} Cepheids (relative to the distance of NGC$\,$4258, 29.4 mag) with one additional offset term for including SNe ($M^0_B$), a total of one linearity term and two offsets. The {\it HST} Cepheid linearity is measured to be $0.994 \pm 0.010$, in good agreement with unity. There is also no evidence of an offset (non-SN) relative to all other indicators, with {\it HST} Cepheids in good accord with the mean. In the inset we show the combined constraint out to 5$\sigma$ confidence and compared to the region that would be necessary to produce an 0.18 mag mean bias between NGC$\,$4258 and the mean {\it HST} SN Ia host, a range of 2.5 mag.}
\label{fg:linearity} 
\end{figure}

  There are 59 independent distance measurements between NGC$\,$4258 and SN Ia hosts in common with the same measured with {\it HST} Cepheids (R22) (we will refer to here as $\mu_{\it HST}$) which can be used to obtain a new constraint on the linearity of the Cepheid distance measurements --- that is, the ratio ${\mu_{\it HST} /\mu_{{\it other}}}$. In Fig.~\ref{fg:linearity} we perform a linear fit, $\mu_{it HST}={\it slope} \times \mu_{\it other}-{\it offset}$ between {\it HST} Cepheid distances. We also can simultaneously include the 37 host measures from SNe~Ia. For SNe we replace $\mu_{\it other}$ with the SN standardized magnitude ($m^0_B$) and allow for a unique SN offset which is $M^0_B$. It is {\it critical} (and we take care) to perform this fit with independent measurements and uncertainties for both axes (i.e., two dimensions and averaging the other methods before comparing to the {\it HST} Cepheid result). We find ${\mu_{\it HST} /\mu_{{\it other}}}=0.994 \pm 0.010$ and an offset of the non-SN data of $-0.01 \pm 0.03$~mag, referenced to $\mu=29.4$~mag, the geometric distance of NGC$\,$4258. The slope matches unity at 0.6$\sigma$ and the offset is consistent with zero. This is the strongest constraint measured for the linearity of {\it HST} Cepheid distances and there is no evidence of a non-linearity. It is also the strongest constraint on an offset between {\it HST} Cepheid distances and all other (non SN) measures, a precision of 0.02 mag. That is, there is no evidence for either a multiplicative or additive bias in the {\it HST} Cepheid distances when compared to all other measures, simultaneously. 
  
\cite{Riess:2022} analyzed the Cepheid distance linearity against only SNe~Ia and found it consistent with unity at 1.5$\sigma$ but this was a weaker constraint due to the lack of the non-SN distances now available. We note that the additional, non-SN~Ia, primary distance indicator data provide a stronger constraint than the SN data alone owing to the smaller errors of the former per object, about half the size or a 4-to-1 weight advantage.

While there is no evidence for a non-linearity, the constraint also strongly out rules a non-linearity as a resolution of the Hubble Tension. A non-linearity would have to produce a change of $5\times \log (73.2/67.5)\sim$0.18 mag between the distance of NGC$\,$4258 and the mean {\it HST}-calibrated SN Ia host ($\mu=29.4$ to $31.9$), a span of 2.5 magnitudes, or a needed bias of 0.07 magnitudes in distance per magnitude which is excluded by the constraint at $\sim$7$\sigma$. A combination of multiplicative and additive {\it HST} Cepheid bias would require $0.18=(2.5\times(1-$slope$)+$offset) which as shown in Fig.~\ref{fg:linearity} is ruled out well beyond the 5$\sigma$ confidence contour.

F24 claimed a 3$\sigma$ correlation by regressing individual {\it HST} Cepheid distances from R22, ($\mu_{\it HST}$), versus SN~Ia $M^0_B=m^0_B-\mu_{\it HST}$ (i.e., included in the above), however, this correlation as appears to be invalid as it includes the same (i.e., fully correlated) variable, $\mu_{\it HST}$, in both the dependent and independent axis (which is used to determine a SN $M^0_B$). We discuss this in Appendix B.
\clearpage
\section{Measuring H$_0$}
\subsection{Expected Differences in H$_0$ from {\it JWST} Subsamples}
\begin{figure}[t!] 
\begin{center}
\includegraphics[width=1.05\textwidth]{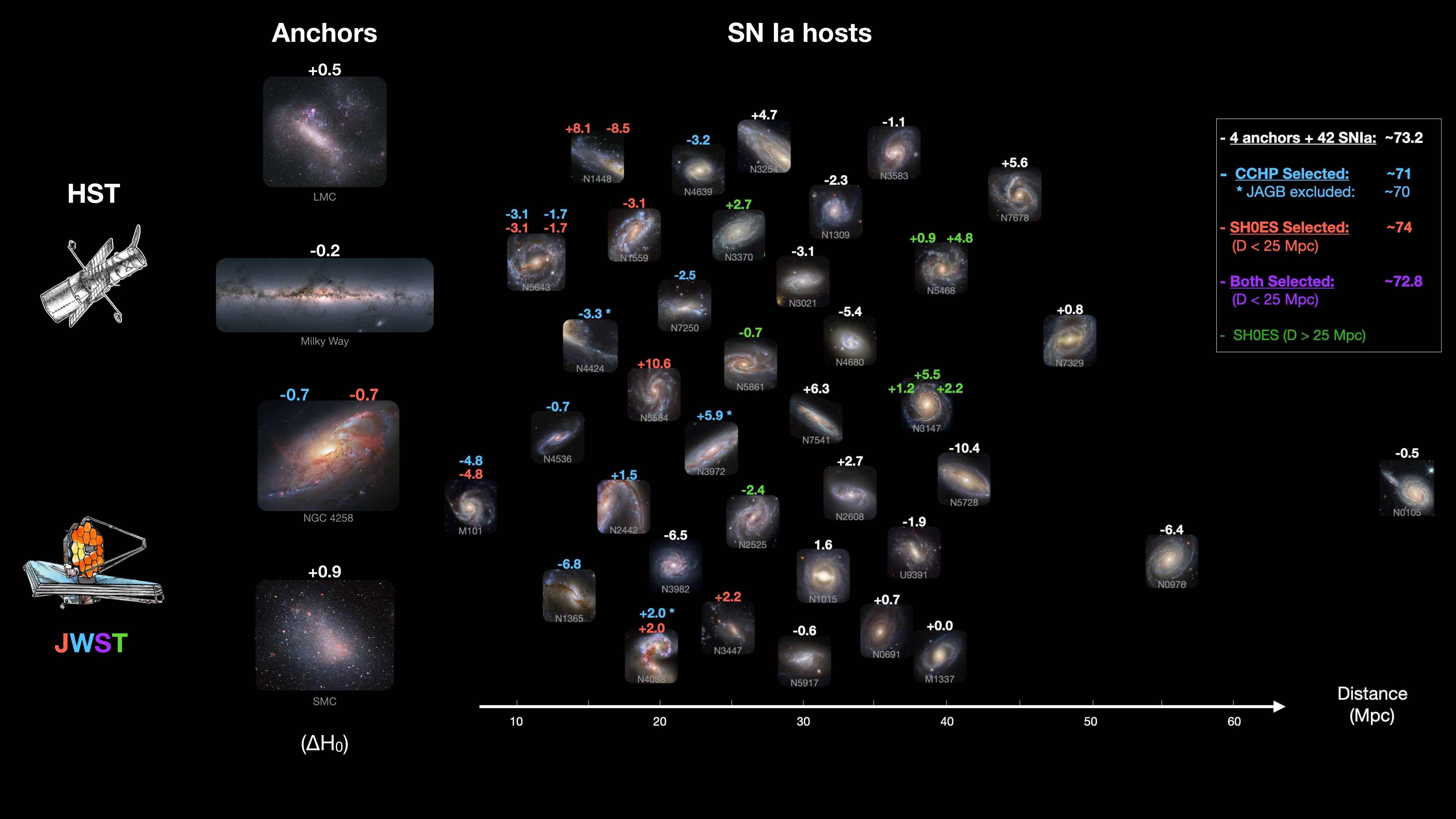}
\end{center}
\caption{Anchors and SN~Ia hosts selected to cross-check {\it HST} and {\it JWST} distances from the full {\it HST} sample of 4 anchors and 42 SNe~Ia. We show the value of H$_0$ indicated by {\it HST} alone for each SN~Ia. Small samples will produce large fluctuations in the value of H$_0$.
({\it HST} also indicates the selection by both teams of the anchor NGC$\,$4258 will produce a lower value of H$_0$ by $-0.7$\kmsmpc.)   The {\it JWST} subsample selected by each team may be directly compared to the same from {\it HST} (i.e., ``apples to apples'') without bias, but if used alone to determine H$_0$ the subsample value is seen to be biased with respect to the full {\it HST} sample mean with the values calculated from {\it HST} Cepheids indicated on the right. A larger sample, both teams combined (red and blue, $N=15$ SNe~Ia), is found to be minimally biased, and nearly complete in distance to $D < 25$~Mpc.}
\label{fg:selections} 
\end{figure}

The measurement of H$_0$ from {\it JWST} host sub-samples alone will show  relatively large variations, due to the small number of SNe they contain, coupled to the substantial SN~Ia intrinsic scatter. The selection differences and their impacts are illustrated in Fig.~\ref{fg:selections}, where we show the difference in H$_0$ we would expect based on the full sample of host {\it HST} Cepheid measurements in R22 and the variation introduced by the anchor. These differences due to selection are often referred to as ``subsample bias," because they measure how the subsample represents the full sample. The driver of this variation is the inherent scatter in the luminosity of individual SNe Ia,\footnote{These variations are known to be intrinsic to SNe~Ia, as they are seen not only in comparison with the Hubble flow, but also in ``siblings'' (multiple SNe in the same host, so independent of distance to the host; \citealp{Scolnic20}).} which is a factor of 2-3 times larger than the uncertainty in the distance estimate to a typical host. For example, the 10 SN Ia calibrated with {\it JWST} in the CCHP sample and analyzed with the Carnegie Supernova Program (CSP) data, where the intrinsic SN Ia dispersion is found to be 0.19 mag \citep{Uddin:2023},  produces a $1\sigma$ uncertainty in H$_0$ of 2.0 \kmsmpc. 

Another difference in $H_0$ between a {\it JWST}-only measurement and the full {\it HST} samples results from the availability of anchors. NGC$\,$4258, the sole anchor available for JWST, produces the lowest value of H$_0$ of the 4 anchors, decreasing H$_0$ from the 4-anchor mean of \citet{Breuval:2024} by $\Delta H_0 \approx 0.7$~\kmsmpc. 

Determining H$_0$ requires the standardized apparent magnitudes of the SN~Ia ($m_B^0$) in the relevant hosts and absolute magnitudes ($M_B^0$) determined as $M_B^0=m_B^0-\mu_0$. For a well-measured mean $M_B^0$, H$_0$ may be determined from 
\begin{equation}
5\, \textrm{log}({\rm H}_0/\MBh)=M_B^0-(\MB)\, .
\end{equation}
This is a useful approximation to the empirical calibration for the Pantheon+ SNe~Ia sample from \citetalias{Riess:2022} (this simplified approximation rather than the full, simultaneous distance-ladder fit including covariance is accurate to $\sim$0.01 mag). Calibrating all 42 SNe~Ia with {\it HST} Cepheids and the single anchor, NGC$\,$4258, produces a mean $M_B = -19.28$~mag and H$_0=72.5\pm 1.5$~\kmsmpc, matching fit \#10 given by \citetalias{Riess:2022}.

\begin{deluxetable*}{lrr}[t!]
\tabletypesize{\small}
\tablewidth{0pt}
\tablecaption{H$_0$ Error Budget for $\sim$10 SN~Ia Measured with {\it JWST} by 3 Methods Calibrated by NGC$\,$4258\label{table:h0err2}}
\tablehead{
\multicolumn{1}{l}{Term} & \colhead{$\sigma(\textrm{stat})$} & \colhead{\hspace*{0.75in}$\sigma(\textrm{sys})$}\\[-6pt]
& \multicolumn{2}{c}{[mag]}}
\startdata
\hline
Cepheid Subtotal (see Table \ref{tab:comperr})& 0.04 & 0.03 \\
TRGB Subtotal (see Table \ref{tab:comperr}) & 0.045 & 0.01--0.08 \\
JAGB Subtotal (see Table \ref{tab:comperr}) & 0.055 & 0.01--0.10 \\
\hline
\textbf{Combining Three Methods} (c,d) & 0.02 & $\sim$0.02 \\
\hline
\hline
\multicolumn{3}{l}{\textbf{Common Uncertainties, Independent of Distance Method}} \\
 \hline
1st rung: Geometric Distance to  NGC$\,$4258 & \multicolumn{1}{c}{--} & 0.032 \\
2nd rung: SN distances to hosts,$^a$ (0.13--0.17)$^e \times N_\textrm{hosts}^{-1/2}$\hspace*{1in}  & 0.043--0.056 & \multicolumn{1}{c}{--} \\
3rd rung: SNe~Ia in Hubble Flow & \multicolumn{1}{c}{--} & 0.01 \\
\hline
\textbf{Common Uncertainty Subtotal} & 0.047--0.059 & 0.04 \\
\hline
\hline
\textbf{Cepheids and Common Total} & 0.068--0.075 & 0.044 \\
\textbf{TRGB and Common Total} & 0.065--0.072 & 0.044 \\
\textbf{JAGB and Common Total} & 0.065--0.072 & 0.044 \\
\textbf{3 Methods and Common Total} & 0.052--0.063 & 0.044 \\
\hline
\hline
\multicolumn{3}{l}{\textbf{3 method stat+sys error in H$_0$: 0.062--0.072 mag or 2.0--2.3~\kmsmpc;}} \\
\multicolumn{3}{l}{\textbf{individual methods $\sim$2.5~\kmsmpc}}\\
\hline
\hline
\enddata
\tablecomments{An uncertainty budget for the present analysis. Uncertainties are listed separately for each distance method (Cepheids, TRGB, JAGB) and their individual subtotals are given. We derive the uncertainties when combining the three in the ``Averaging between methods." A separate list of uncertainties, which are common to each distance method (e.g., the uncertainties from  SN measurements) are also given. Finally, we combine the Averaged Uncertainty with the Common Uncertainty to derive a total H$_0$ uncertainty. (a) Pantheon + SN sample \cite{Brout:2022} has a mean per SN~Ia error of 0.13 mag, CSP SN sample \citep{Uddin:2023} finds $\sigma_{\rm int} = 0.17$ per SN for CSP I \& II.}
\end{deluxetable*}
We now predict differences in H$_0$ {\it due only to the selection of a {\it JWST} subsample} by comparing the $M_B^0$ subsample means {\it determined from the {\it HST} Cepheid distances} using equation 6. In Table~\ref{tab:distpar} we list the relevant quantities for the full sample of 42 SNe~Ia from \citetalias{Riess:2022} as well as the SN~Ia host subsamples selected for {\it JWST} studies. The comparisons between expected (HST) and actual (JWST) measure of $H_0$ are shown in Fig.~\ref{fg:compareh0}.

\subsection{CCHP {\it JWST} Sample Difference}

For the 10 hosts of 11 SNe~Ia selected by the CCHP for {\it JWST} observations, {\it HST} Cepheids gave $M_B^0=-19.32\pm0.05$~mag (where the uncertainty includes the measurement of NGC$\,$4258, $\pm0.04$ without it) and a corresponding value of H$_0=71.2$~\kmsmpc, thus 2.0 \kmsmpc~lower than the full {\it HST} sample---simply due to the choice of SN hosts and anchor. 

An even larger sample difference is expected for the smaller subsample of 7 SN~Ia hosts for which the CCHP {\it JWST} JAGB measurements are provided \citep{Lee:2024}. {\it A priori}, {\it HST} Cepheid measurements (see Table 1) find $M_B=-19.34\pm0.05$~mag and H$_0$=70.3~\kmsmpc\ for this JAGB subset, {\it an expected reduction of 3.0 \kmsmpc from the full {\it HST} sample due to selection}. The use of CSP instead of the Pantheon$+$ SN samples reduces this by another 0.7 \kmsmpc~according to L24, and the selection of the mode statistic rather than the mean or median reduces $H_0$ by another 1.1 \kmsmpc. The CCHP JAGB analysis excludes 3 SN~Ia hosts (NGC$\,$3972, 4038, and 4424) from the CCHP sample of 10, and these 3 excluded objects have a higher expected average, H$_0=75$~\kmsmpc, based on their {\it HST} Cepheid measurements; among the three excluded objects are the two with the highest H$_0$ from the CCHP sample (NGC$\,$3972 and 4038). 

The subsample chosen for the CCHP JAGB measurements produces an unusual combination of draws from the full {\it HST} sample, being both low in H$_0$ and in a very tight cluster, each a $\sim$2$\sigma$ level fluctuation. The SN~Ia dispersion in $M_B$ of the set is uncommonly low at 0.06~mag, less than half the typical SN~Ia population dispersion \citep{Brout:2022,Scolnic:2022}. As shown in Fig.~\ref{fg:histograms}, only $\sim$5\% of randomly-selected samples of 8 SNe~Ia from the 42 yield an {\it HST} Cepheid predicted value of H$_0$ this low, and $\sim$5\% of samples have SNe with such a small variation in luminosity. These subsample characteristics seen from {\it HST} Cepheid measurements are well matched in the actual {\it JWST} JAGB measurements (L24). Because the subsample is unusual {\it a priori}, a standard uncertainty propagated to H$_0$ will not account for the uncommonly high difference between this subsample and the full sample mean. This illustrates the greater value of comparing distance measures with one method to distance measures with another for subsamples rather than comparing H$_0$ or SN scatter which avoids the issue of uncommon or unrepresentative subsamples. 
 
 The value of H$_0$ found by F24, which is lower than that of R22, is anticipated by the difference of the selected subsample relative to the {\it HST} full sample mean and the exclusion of three galaxies for JAGB measurements (see Fig.~\ref{fg:comparedis}, vertical blue line). F24 report {\it JWST} Cepheids in the CCHP-selected sample give a value of $\sim$72~\kmsmpc\ and NIR-TRGB $\sim$70~\kmsmpc, both near the {\it HST}-expected $\sim$71~\kmsmpc\ for the subsample, and {\it JWST} JAGB gives $\sim$69~\kmsmpc, near the {\it HST} measured for that subsample (with three excluded) of 70.3~\kmsmpc; see Fig.~\ref{fg:histograms}, all with the Pantheon$+$ SN sample. For the combination of these three methods, F24 reports $70$~\kmsmpc, near the {\it HST}-expected value of $\sim$70.8$\pm 2.3$~\kmsmpc. In summary, the difference of this selected SN subsample relative to the mean and the exclusion of several hosts for JAGB measurements reduce H$_0$ as expected by the full sample. In Fig.~\ref{fg:selections} we show the individual SN~Ia values of H$_0$ for the full {\it HST} sample and the differences from anchor choice as a full summary.
 
 For the SH0ES {\it JWST} host selection of 11 SNe~Ia in 8 hosts, {\it HST} Cepheids predict $M_B^0=-19.25\pm0.05$~mag and H$_0=73.9 \pm 2.3$~\kmsmpc\ (or $-19.26$~mag and 73.6~\kmsmpc\ for 9 SNe~Ia with $D\leq 25$~Mpc). This represents a $\sim$1$\sigma$ fluctuation due to the SN selection in the other direction (a higher H$_0$). This compares well with the {\it JWST} Cepheid measurement of $H_0=74.2$ \kmsmpc.

\begin{figure}[b!] 
\begin{center}
\includegraphics[width=0.7\textwidth]{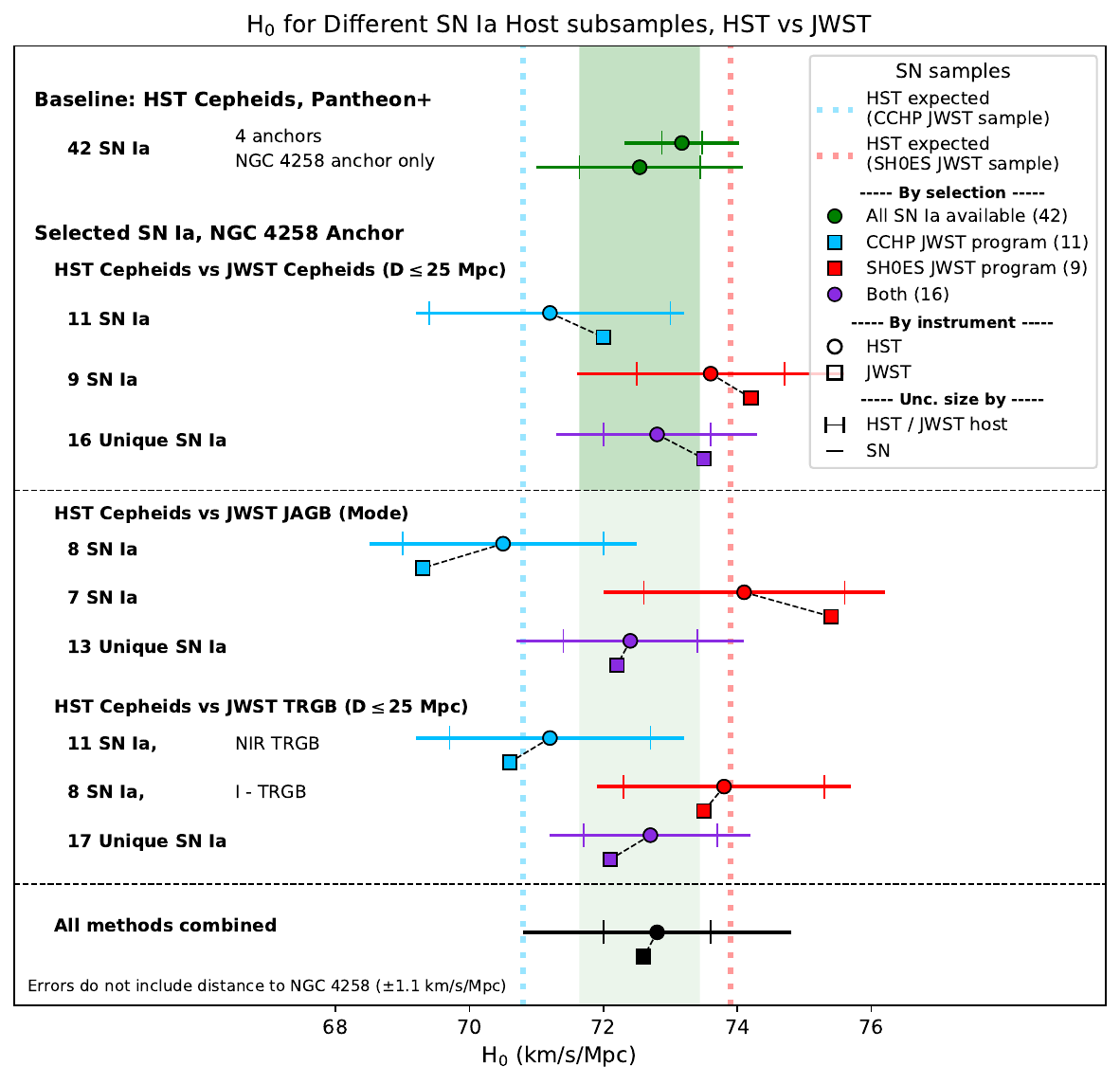}
\end{center}
\caption{Comparisons of H$_0$ between {\it HST} Cepheids and
other measures ({\it JWST} Cepheids, {\it JWST} JAGB, and {\it JWST} NIR-TRGB) for SN~Ia host subsamples selected by different teams and for the different methods. The top section shows the values for H$_0$ using 4 geometric anchors, and also using only NGC$\,$4258. Below, for each selected subsample (by Team or method), we show the value of H$_0$ based on the {\it HST} Cepheid measurements (from R22) and from the {\it JWST} distance measurements. The smaller capped error bars indicate the uncertainty from the distance measures between the first/second rung distance-measure (Cepheids/TRGB/JAGB) alone and the full error bar includes the SN data. The CCHP and SH0ES subsamples selected for {\it JWST} observations produce a large difference of 3--4~\kmsmpc\ in H$_0$ {\it a priori} owing to selection. The {\it HST} Cepheid and {\it JWST} distance measurements themselves are in good agreement. }
\label{fg:compareh0} 
\end{figure}

 The full error bars in Fig.~\ref{fg:compareh0} include the noise from the small subsample of SN measurements. Thus for a new measurement with {\it JWST} in comparison to {\it HST}, the differential errors as given in Table 2, can range from 0.03$-$0.10~mag or 1$-$3~\kmsmpc. 
\clearpage
\begin{deluxetable*}{lllrrr}[t!]
\tabletypesize{\small}
\tablewidth{0pt}
\tablecaption{Comparison of Measured and Expected Values of H$_0$ (with Pantheon+ SN)\label{table:H0_comparisons}}
\tablehead{\multicolumn{2}{c}{Sample} & \multicolumn{1}{c}{Telescope} & \multicolumn{1}{c}{\ \ H$_0$\ \ } & \multicolumn{2}{c}{$\sigma($H$_0)$}\\
 & & & & \colhead{Total} & \colhead{\it HST/} \\[-6pt]
 & & & & & \colhead{\it JWST}}
\startdata
\multicolumn{6}{c}{\bf Baseline: {\it HST} Cepheids, Pantheon+} \\
\hline
\textbf{42 SN~Ia} & \textbf{4 Anchors}          & {\it HST} & \textbf{73.17} & \textbf{0.86} & 0.30  \\
\textbf{42 SN~Ia} & \textbf{NGC$\,$4258 Anchor\hspace*{0.5in}} & \textbf{{\it HST}} & \textbf{72.54} & \textbf{1.54} & 0.90  \\
\hline
\multicolumn {6}{c}{\bf{\it JWST} Cepheids Measured \& {\it HST} Cepheid Expected}\\
\hline
11 SN~Ia & CCHP Selected      & {\it HST} & 71.2  &       &      \\
         &                    & {\it JWST}& 72.0  & 2.4   & 1.8  \\
9 SN~Ia  & SH0ES Selected     & {\it HST} & 73.6  &       &      \\
         & ($D\leq 25$ Mpc)       & {\it JWST}& 74.2  & 2.3   & 1.1  \\
16 SN~Ia &  Both  Unique SN     & {\it HST} & 72.9  &       &      \\
         & ($D\leq 25$ Mpc)       & {\it JWST}& 73.4  & 2.1   & 1.1  \\
\hline
\multicolumn {6}{c}{\bf{\it JWST} JAGB (Mode) \& {\it HST} Cepheid Expected} \\
\hline
8 SN~Ia & CCHP Selected  & {\it HST}  & 70.3 &   &    \\
        &    mode            & {\it JWST} & 68.9 & 2.4 & 1.5  \\
        &    mean/median       & {\it JWST} & 70.0 & 2.4 & 1.5  \\
7 SN~Ia & SH0ES Selected & {\it HST}  & 74.1 &   &    \\
        &                & {\it JWST} & 75.4 & 3.1 & 1.5  \\
13 SN~Ia & Both Unique  SN       & {\it HST}  & 72.4  &  &    \\
         &               & {\it JWST} & 72.2  &  2.3    & 1.3      \\
\hline
\multicolumn {6}{c}{\bf{\it JWST} NIR-TRGB \& {\it HST} Cepheid Expected} \\
\hline
11 SN~Ia &  CCHP Selected & {\it HST}  & 71.2 &   &    \\
         &                & {\it JWST} & 70.1 & 2.4 & 1.5  \\
\hline
\multicolumn {6}{c}{\bf{\it JWST} I-TRGB ($D \leq$ 25 Mpc) \& {\it HST} Cepheid Expected} \\
\hline
8 SN~Ia & SH0ES Selected & {\it HST}  & 73.8 &   &    \\
        &                & {\it JWST} & 73.5 &  2.4    &   1.5  \\
        \hline
17 SN~Ia & Both  Unique SN       & {\it HST}  & 72.7  &    &    \\
         &               & {\it JWST} & 72.1  & 2.2  & 1.2      \\
\hline
\multicolumn {6}{c}{\bf{\it JWST} All Methods ($D \leq$ 25 Mpc) \& {\it HST} Cepheid Expected} \\
\hline
\textbf{13-17 SN~Ia\hspace*{0.5in}} & \textbf{All} & {\it \textbf{HST}}  & \textbf{72.8} &   &    \\
        &                & {\it \textbf{JWST}} & \textbf{72.6} &  \textbf{2.0}    &  1.0   \\
\hline
\enddata
\tablecomments{Subsample uncertainties do not include 1.1~\kmsmpc\ error from maser distance to NGC$\,$4258.}
\end{deluxetable*}

\subsection{H$_0$ from Joint Samples}

To avoid a selection bias relative to the SN~Ia population, it is important to define a (combined) sample which is complete to some distance (a common method to avoid magnitude bias and also to avoid the vagaries of human selection). We note that a near-complete sample of SN Ia (in R22) to $D \leq 25$ Mpc with {\it JWST} observations is formed by combining both the above SH0ES and CCHP samples for a total of 14 (unique) hosts of 16 SNe~Ia, missing only SN$\,$1998aq (NGC$\,$3982) $D \leq 25$ Mpc which neither team targeted. A combined $D \leq 25$ Mpc {\it JWST} sample excludes one host of two SNe~Ia,  NGC$\,$5468 from the SH0ES {\it JWST} sample which is at $D=40$ Mpc. The CCHP sample of 10 alone does not include hosts of 6 SNe~Ia at distances $\leq$ 25 Mpc or at distances nearer than their most distant (NGC$\,$4639, $\mu \leq 32.0$ at 95\% confidence): SN$\,$2005df, SN$\,$2007af, SN$\,$2001el, SN$\,$2012ht, SN$\,$2021pit, and SN$\,$1998aq. The merged {\it JWST} $D \leq 25$ Mpc sample of 16 SN Ia is also seen to be more representative of the full {\it HST} sample of 42 SN Ia than either group's selected sub-sample, an expected consequence of ``reversion to the mean'' as the samples grow. {\it HST} Cepheids predict H$_0=72.9 \pm 2.1$~\kmsmpc\ for the {\it JWST} $D \leq 25$ Mpc sample. 

For the merged $D \leq$ 25~Mpc sample of 16 SNe~Ia, we find from {\it JWST} Cepheids  H$_0=73.4\pm2.0$~\kmsmpc, similar to the {\it HST} result from this same sample.
The joint sample of {\it JWST} JAGB and TRGB measurements yield H$_0=72.2\pm2.3$~\kmsmpc\ and H$_0=72.1\pm2.3$~\kmsmpc, respectively. The {\it HST} Cepheid expectation for the JAGB $H_0$ is somewhat smaller at 72.4 \kmsmpc\ due to the exclusion of several hosts in L24. Finally, we can combine all three methods, which yields H$_0=72.6\pm2.0$~\kmsmpc, in good agreement with the value expected from {\it HST} Cepheids for the same sample of H$_0=72.8\pm2.0$~\kmsmpc. The uncertainties for this combination are explained in Table~\ref{table:h0err2}. For these estimates we used the minimum systematic error listed in Table~1 for each method.  For an expanded JWST sample of SN Ia hosts with smaller statistical uncertainty we would advocate a more comprehensive analysis of systematic uncertainties.
When we compare the {\it HST} predicted values of H$_0$ to those found with {\it JWST} measurements, we find the values inferred using {\it JWST} are consistent with expectations for every subsample. We can see that the measured values are at or within $1\sigma$ of these smaller uncertainties, i.e., between {\it HST} and {\it JWST} excluding SNe, for each comparison set. We reiterate that the {\it HST} expectations above all come from the use of NGC 4258 as a single anchor, yielding lower values of $H_0$ by 0.5 than the three anchor calibrations in R22.

\begin{figure}[t!] 
\begin{center}
\includegraphics[width=0.85\textwidth]{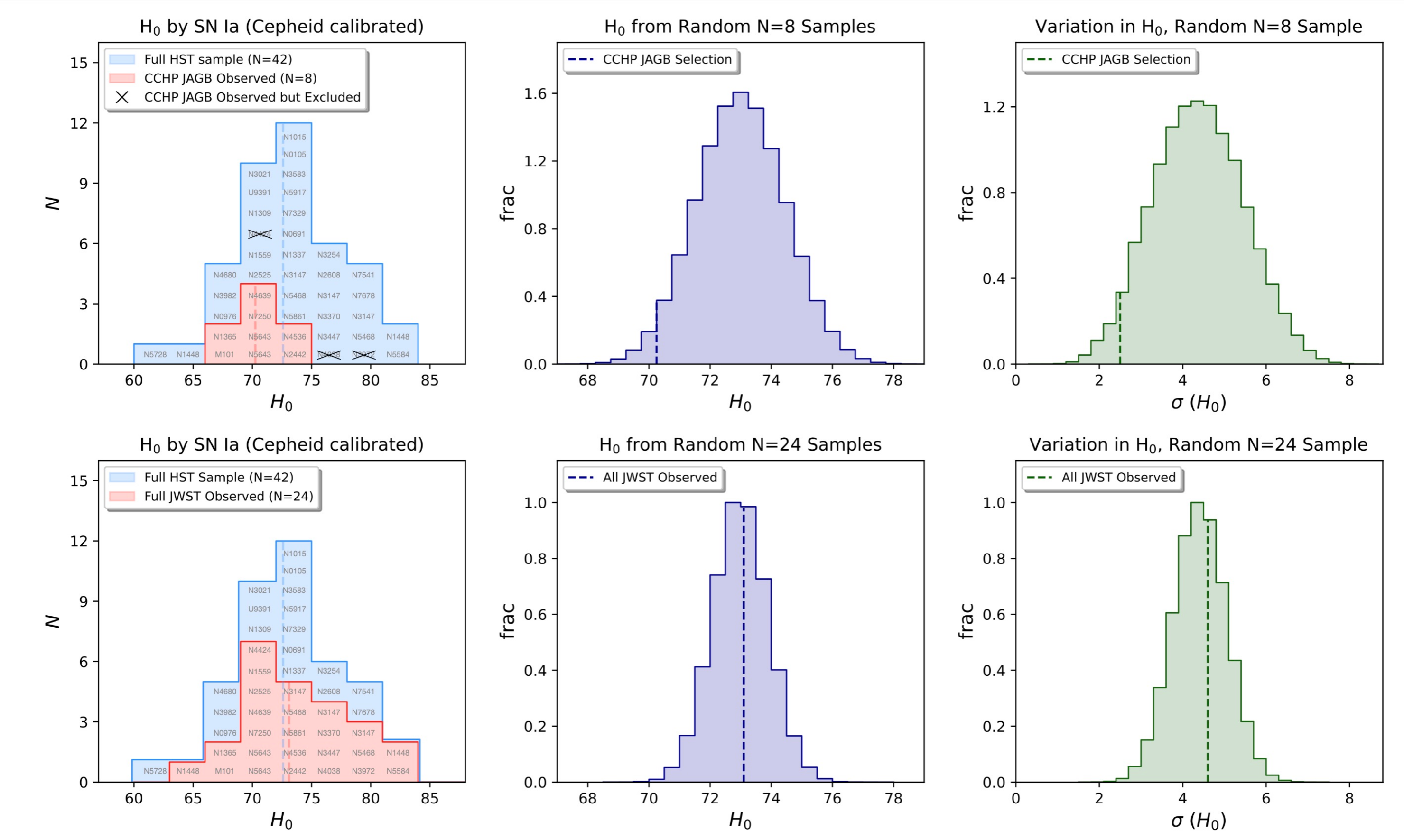}
\end{center}
\caption{Distribution of H$_0$ values for each calibrator SN~Ia as calibrated by {\it HST} Cepheids (left panel). The mean of the complete set of 42, in blue, results in H$_0=73$~\kmsmpc, and the standard deviation is a result of SN~Ia dispersion with $\sigma$ of 6\% in distance. We show the subsample selected by the CCHP (10 hosts of 11 SNe~Ia) and after their exclusion of 3 hosts for JAGB measurements. The remaining set of 8 SNe~Ia used for JAGB, in red, are biased low with respect to the mean, with {\it HST} Cepheids expecting H$_0=70.5$~\kmsmpc\ for this set. Selecting every unique combination of 8 SNe~Ia from the original 42 shows the selected JAGB sample to be unusual, both low in H$_0$ (middle panel, only 5\% are lower) and with little variation in H$_0$ (right panel, only 5\% are lower), with $< 0.1$\% of samples both lower and tighter. This selection is the primary reason for the low value of H$_0$ from CCHP JAGB, not from a difference in measured distances.}
\label{fg:histograms} 
\end{figure}

For the merged sample, the averaging of distance methods occurs before the use of SN data to avoid double-counting\footnote{The small quoted errors of $\pm 1$ for $H_0$ in F24 via their Bayesian route appear to be a consequence of directly multiplying the H$_0$ likelihood for each method. Since each method calibrates the very same set of SNe, this treats 10 SNe~Ia as equivalent to the statistical power of 30 and therefore underestimates the uncertainty in H$_0$ by $\sim\sqrt{3}$. See Appendix B for further discussion.}. It is interesting to examine the much larger sample of all SN~Ia hosts observed to date with {\it JWST} which includes 24 of the 42 {\it HST} SNe~Ia. The full SN sample with {\it JWST} observations (to date) is highly representative of the {\it HST} SN sample as shown in Fig.~\ref{fg:histograms} in terms of expected H$_0$ and variance with little sample bias. When increasing the number of SN hosts, we can see from the widths of the middle panels in Fig.~\ref{fg:histograms} how the second-rung uncertainty from SNe will decrease. The impact of selection when comparing {\it HST} and {\it JWST} will diminish with the growing samples.
\section{Discussion \label{sec:discussion}}

\begin{figure}[b!] 
\begin{center}
\includegraphics[width=0.53\textwidth]{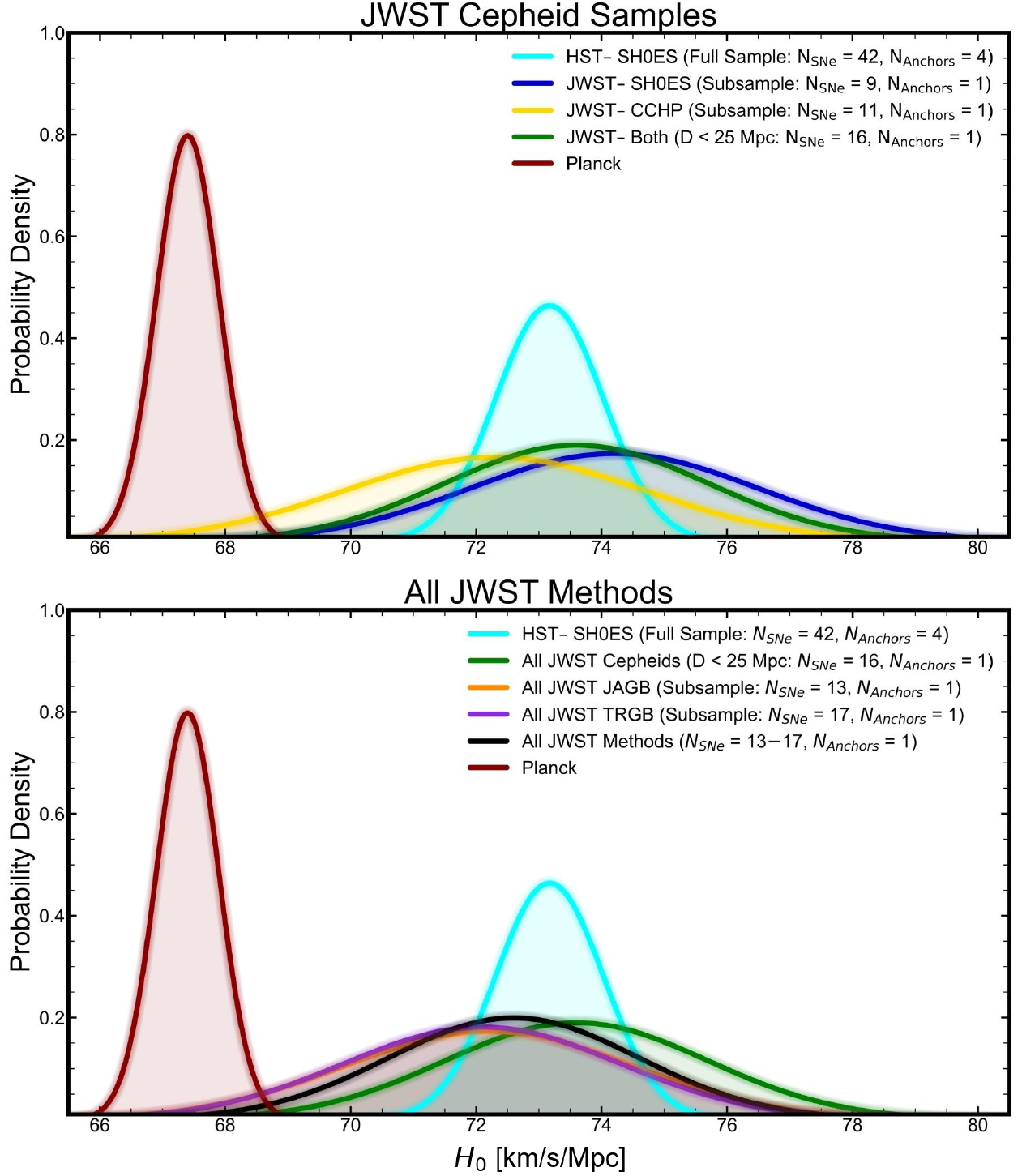}
\end{center}
\caption{Comparisons of H$_0$ between {\it HST} Cepheids and
other subsamples of {\it JWST} Cepheids and anchors. }
\label{fg:gaussians} 
\end{figure}

A complete {\it JWST} ladder, still ``under construction,'' is necessarily weaker than the one built by {\it HST} over decades as it is limited to a single anchor rather than 4 \citep{Breuval:2024}, resulting in less precision, loss of redundancy, and a reduction in resolution of the tension at inception. Limiting the {\it HST} ladder to the same single anchor as {\it JWST}, NGC$\,$4258, results in H$_0=72.5\pm1.5$~\kmsmpc\ \citepalias{Riess:2022}, a reduction of the tension to $3\sigma$ before any comparison with {\it JWST}. Limiting to one anchor and the 17 SNe~Ia at $D< 25$ Mpc trivially further reduces the significance of the tension, H$_0=72.3\pm1.8$~\kmsmpc, but only masks the tension rather than offering any explanation.   The situation is illustrated in Fig.~\ref{fg:gaussians}.  While {\it JWST} is enormously powerful for checking {\it HST} distances, it only weakly constrains the tension due to its lack of statistics from SNe and anchors.

Simply summarized, {\it JWST} offers the means to test {\it HST} on the second rung, but given their demonstrated consistency, there is no reason not to use the full {\it HST} ladder --- its SN sample is complete, it has superior statistics, and it uses multiple, redundant anchors to measure H$_0$. However, the expanding {\it JWST} sample will gradually remedy the disparity. We show here that a combination of all {\it JWST} subsamples is already nearly complete to $D\leq 25$~Mpc, reaches 40\% of the {\it HST} SN sample, and reduces the {\it a priori} bias, an expected consequence of reversion to the mean of larger samples. 

Finally, we remark that the present circumstance is not unusual in observational cosmology; a comparable situation was seen for CMB measurements. The South Pole Telescope (SPT) has measured CMB fluctuations on finer angular scales than {\it Planck} but over smaller patches of the sky, including 2500 square degrees in the initial SPT-SZ survey \citep{Story:2013}, and 500 square degrees for the SPTpol survey \citep{Henning:2018}. Fitting the $\Lambda$CDM model to these data sets the SPT team found H$_0=75.0 \pm 3.5$ and $71.3 \pm 2.1$~\kmsmpc, respectively, higher than the {\it Planck} constraints and in good agreement with the SH0ES distance ladder. However, where the SPT and {\it Planck} measurements were directly comparable (i.e., on the SPT patch, for the range of angular scales accessible to both instruments), they were found to agree well with one another, with a null-test PTE of 0.30 \citep{Hou:2018}. The conclusion, therefore, was not that the SPT data indicated a problem with the full-sky {\it Planck} data or vice-versa, but that the SPT data provided a valuable new consistency check of the {\it Planck} measurements, which was passed, and that the SPT H$_0$ results differed from the {\it Planck} results because of sample variance on that patch of sky (i.e., it was not exactly representative of the full sky). 

Future data might help resolve the Hubble tension problem, but not simply by yielding a high or low value of $H_0$ (or even an intermediate result). Instead, we think it will be necessary to understand why different types of measurements have yielded inconsistent results for a decade. If there is an observational or data analysis error with the CMB or local distance ladder, it is far from clear what it might be. Studies that provide consistency tests are an important part of seeking a solution. Measurements from two different CMB space missions agree \citep{Addison:2018}, ruling out a number of potential candidate causes of an erroneous Hubble constant. The community-wide effort to seek cosmological solutions to the tension are valuable, but so far have largely demonstrated the difficulty in explaining the tension. Whatever the explanation it seems fair to say that it is elusive and that we, as a community, are missing something.

\section{Conclusion \label{sec:conclusion}}

A major success of the first years of {\it JWST} has been its ability to provide a number of cross-checks on the local distance ladder as measured using {\it HST}. In this paper, we surveyed measurements using {\it HST} and {\it JWST} with multiple techniques including Cepheids, TRGB, and JAGB to search for any systematic biases in these measurements. We find that {\it HST} Cepheid measurements are consistent at the 0.03 mag and $1\sigma$ level with all other techniques from the two telescopes. By comparing all distance indicators to common hosts, {\it HST} Cepheids versus others, we find the strongest constraint to date on the linearity in the {\it HST} Cepheid measurements,  $0.994\pm0.010$, with no significant evidence of non-linearity, and more than 5$\sigma$ from either a multiplicative or additive bias needed to resolve the Hubble Tension. We show that different values found for H$_0$ based solely on {\it JWST} can be traced to differences in the small samples of SN hosts (and their SNe) and anchors selected for early {\it JWST} observations. When combining all the {\it JWST} measurements for each technique, we find  $73.4\pm2.1$, $72.2\pm2.2$, and $72.1\pm2.2$~\kmsmpc\ for {\it JWST} Cepheids, JAGB, and TRGB, respectively. When we combine all the methods (but each SN measurement included only once), we determine H$_0=72.6\pm2.0$~\kmsmpc, in good agreement with 72.8\kmsmpc\ that {\it HST} Cepheids would yield for the same sample. While it will still take multiple years for the {\it JWST} sample of SN hosts to be as large as the {\it HST} sample, we show that the current {\it JWST} measurements have already ruled out systematic biases from the first rungs of the distance ladder at a much smaller level than the Hubble tension. 

\section{Acknowledgments \label{sec:acknowledgments}}

We are indebted to all of those who spent years and even decades bringing {\it JWST} to fruition. This research has made use of NASA’s Astrophysics Data System. D.S. is supported by Department of Energy grant DE-SC0010007, the David and Lucile Packard Foundation, the Templeton Foundation and Sloan Foundation. G.S.A acknowledges financial support from {\it JWST} GO–1685 and GO-2875. C.D.H. acknowledges financial support from {\it HST} GO-16744 and GO-17312. A.V.F. is grateful for support from the Christopher R. Redlich Fund and many other donors. 

If something does not make sense, please email ariess@stsci.edu with questions.

\bibliographystyle{aasjournal}
\bibliography{bibsh0es}

\begin{thebibliography}{}
\expandafter\ifx\csname natexlab\endcsname\relax\def\natexlab#1{#1}\fi
\providecommand{\url}[1]{\href{#1}{#1}}
\providecommand{\dodoi}[1]{doi:~\href{http://doi.org/#1}{\nolinkurl{#1}}}
\providecommand{\doeprint}[1]{\href{http://ascl.net/#1}{\nolinkurl{http://ascl.net/#1}}}
\providecommand{\doarXiv}[1]{\href{https://arxiv.org/abs/#1}{\nolinkurl{https://arxiv.org/abs/#1}}}

\bibitem[{{Addison} {et~al.}(2018){Addison}, {Watts}, {Bennett}, {Halpern}, {Hinshaw}, \& {Weiland}}]{Addison:2018}
{Addison}, G.~E., {Watts}, D.~J., {Bennett}, C.~L., {et~al.} 2018, \apj, 853, 119, \dodoi{10.3847/1538-4357/aaa1ed}

\bibitem[{{Anand} {et~al.}(2022){Anand}, {Tully}, {Rizzi}, {Riess}, \& {Yuan}}]{Anand:2022}
{Anand}, G.~S., {Tully}, R.~B., {Rizzi}, L., {Riess}, A.~G., \& {Yuan}, W. 2022, \apj, 932, 15, \dodoi{10.3847/1538-4357/ac68df}

\bibitem[{{Anand} {et~al.}(2024){Anand}, {Riess}, {Yuan}, {Beaton}, {Casertano}, {Li}, {Makarov}, {Makarova}, {Tully}, {Anderson}, {Breuval}, {Dolphin}, {Karachentsev}, {Macri}, \& {Scolnic}}]{Anand:2024}
{Anand}, G.~S., {Riess}, A.~G., {Yuan}, W., {et~al.} 2024, \apj, 966, 89, \dodoi{10.3847/1538-4357/ad2e0a}

\bibitem[{{Anderson} {et~al.}(2024){Anderson}, {Koblischke}, \& {Eyer}}]{Anderson:2024}
{Anderson}, R.~I., {Koblischke}, N.~W., \& {Eyer}, L. 2024, \apjl, 963, L43, \dodoi{10.3847/2041-8213/ad284d}

\bibitem[{{Beaton} {et~al.}(2018){Beaton}, {Bono}, {Braga}, {Dall'Ora}, {Fiorentino}, {Jang}, {Mart{\'\i}nez-V{\'a}zquez}, {Matsunaga}, {Monelli}, {Neeley}, \& {Salaris}}]{Beaton:2018}
{Beaton}, R.~L., {Bono}, G., {Braga}, V.~F., {et~al.} 2018, \ssr, 214, 113, \dodoi{10.1007/s11214-018-0542-1}

\bibitem[{{Breuval} {et~al.}(2024){Breuval}, {Riess}, {Casertano}, {Yuan}, {Macri}, {Romaniello}, {Murakami}, {Scolnic}, {Anand}, \& {Soszy{\'n}ski}}]{Breuval:2024}
{Breuval}, L., {Riess}, A.~G., {Casertano}, S., {et~al.} 2024, arXiv e-prints, arXiv:2404.08038, \dodoi{10.48550/arXiv.2404.08038}

\bibitem[{{Brout} {et~al.}(2022){Brout}, {Scolnic}, {Popovic}, {Riess}, {Carr}, {Zuntz}, {Kessler}, {Davis}, {Hinton}, {Jones}, {Kenworthy}, {Peterson}, {Said}, {Taylor}, {Ali}, {Armstrong}, {Charvu}, {Dwomoh}, {Meldorf}, {Palmese}, {Qu}, {Rose}, {Sanchez}, {Stubbs}, {Vincenzi}, {Wood}, {Brown}, {Chen}, {Chambers}, {Coulter}, {Dai}, {Dimitriadis}, {Filippenko}, {Foley}, {Jha}, {Kelsey}, {Kirshner}, {M{\"o}ller}, {Muir}, {Nadathur}, {Pan}, {Rest}, {Rojas-Bravo}, {Sako}, {Siebert}, {Smith}, {Stahl}, \& {Wiseman}}]{Brout:2022}
{Brout}, D., {Scolnic}, D., {Popovic}, B., {et~al.} 2022, \apj, 938, 110, \dodoi{10.3847/1538-4357/ac8e04}

\bibitem[{{Chandar} {et~al.}(2021){Chandar}, {Bolatto}, {Dale}, {Goudfrooij}, {Klessen}, {Mok}, {Schinnerer}, {Smith}, \& {Walter}}]{2021jwst.prop.2581C}
{Chandar}, R., {Bolatto}, A., {Dale}, D., {et~al.} 2021, {Come Out, Come Out, Wherever You Are: Seeking All the Massive Young Clusters Hidden in the Antennae}, JWST Proposal. Cycle 1, ID. \#2581

\bibitem[{{Dalcanton} {et~al.}(2012){Dalcanton}, {Williams}, {Lang}, {Lauer}, {Kalirai}, {Seth}, {Dolphin}, {Rosenfield}, {Weisz}, {Bell}, {Bianchi}, {Boyer}, {Caldwell}, {Dong}, {Dorman}, {Gilbert}, {Girardi}, {Gogarten}, {Gordon}, {Guhathakurta}, {Hodge}, {Holtzman}, {Johnson}, {Larsen}, {Lewis}, {Melbourne}, {Olsen}, {Rix}, {Rosema}, {Saha}, {Sarajedini}, {Skillman}, \& {Stanek}}]{Dalcanton:2012}
{Dalcanton}, J.~J., {Williams}, B.~F., {Lang}, D., {et~al.} 2012, \apjs, 200, 18, \dodoi{10.1088/0067-0049/200/2/18}

\bibitem[{{Di Valentino} {et~al.}(2021){Di Valentino}, {Mena}, {Pan}, {Visinelli}, {Yang}, {Melchiorri}, {Mota}, {Riess}, \& {Silk}}]{DiValentino:2021}
{Di Valentino}, E., {Mena}, O., {Pan}, S., {et~al.} 2021, Classical and Quantum Gravity, 38, 153001, \dodoi{10.1088/1361-6382/ac086d}

\bibitem[{{Dolphin}(2016)}]{2016ascl.soft08013D}
{Dolphin}, A. 2016, {DOLPHOT: Stellar photometry}, Astrophysics Source Code Library, record ascl:1608.013

\bibitem[{{Durbin} {et~al.}(2020){Durbin}, {Beaton}, {Dalcanton}, {Williams}, \& {Boyer}}]{Durbin:2020}
{Durbin}, M.~J., {Beaton}, R.~L., {Dalcanton}, J.~J., {Williams}, B.~F., \& {Boyer}, M.~L. 2020, \apj, 898, 57, \dodoi{10.3847/1538-4357/ab9cbb}

\bibitem[{{Freedman}(2021)}]{Freedman:2021}
{Freedman}, W.~L. 2021, \apj, 919, 16, \dodoi{10.3847/1538-4357/ac0e95}

\bibitem[{{Freedman} \& {Madore}(2023)}]{Freedman:2023}
{Freedman}, W.~L., \& {Madore}, B.~F. 2023, \jcap, 2023, 050, \dodoi{10.1088/1475-7516/2023/11/050}

\bibitem[{{Freedman} {et~al.}(2021){Freedman}, {Madore}, {Hoyt}, {Jang}, {Lee}, \& {Owens}}]{2021jwst.prop.1995F}
{Freedman}, W.~L., {Madore}, B.~F., {Hoyt}, T., {et~al.} 2021, {Answering the Most Important Problem in Cosmology Today: Is the Tension in the Hubble Constant Real?}, JWST Proposal. Cycle 1, ID. \#1995

\bibitem[{{Freedman} {et~al.}(2024){Freedman}, {Madore}, {Jang}, {Hoyt}, {Lee}, \& {Owens}}]{Freedman:2024}
{Freedman}, W.~L., {Madore}, B.~F., {Jang}, I.~S., {et~al.} 2024, arXiv e-prints, arXiv:2408.06153, \dodoi{10.48550/arXiv.2408.06153}

\bibitem[{{Freedman} {et~al.}(2019){Freedman}, {Madore}, {Hatt}, {Hoyt}, {Jang}, {Beaton}, {Burns}, {Lee}, {Monson}, {Neeley}, {Phillips}, {Rich}, \& {Seibert}}]{Freedman:2019}
{Freedman}, W.~L., {Madore}, B.~F., {Hatt}, D., {et~al.} 2019, \apj, 882, 34, \dodoi{10.3847/1538-4357/ab2f73}

\bibitem[{{Freedman} {et~al.}(2020){Freedman}, {Madore}, {Hoyt}, {Jang}, {Beaton}, {Lee}, {Monson}, {Neeley}, \& {Rich}}]{2020ApJ...891...57F}
{Freedman}, W.~L., {Madore}, B.~F., {Hoyt}, T., {et~al.} 2020, \apj, 891, 57, \dodoi{10.3847/1538-4357/ab7339}

\bibitem[{{Henning} {et~al.}(2018){Henning}, {Sayre}, {Reichardt}, {Ade}, {Anderson}, {Austermann}, {Beall}, {Bender}, {Benson}, {Bleem}, {Carlstrom}, {Chang}, {Chiang}, {Cho}, {Citron}, {Corbett Moran}, {Crawford}, {Crites}, {de Haan}, {Dobbs}, {Everett}, {Gallicchio}, {George}, {Gilbert}, {Halverson}, {Harrington}, {Hilton}, {Holder}, {Holzapfel}, {Hoover}, {Hou}, {Hrubes}, {Huang}, {Hubmayr}, {Irwin}, {Keisler}, {Knox}, {Lee}, {Leitch}, {Li}, {Lowitz}, {Manzotti}, {McMahon}, {Meyer}, {Mocanu}, {Montgomery}, {Nadolski}, {Natoli}, {Nibarger}, {Novosad}, {Padin}, {Pryke}, {Ruhl}, {Saliwanchik}, {Schaffer}, {Sievers}, {Smecher}, {Stark}, {Story}, {Tucker}, {Vanderlinde}, {Veach}, {Vieira}, {Wang}, {Whitehorn}, {Wu}, \& {Yefremenko}}]{Henning:2018}
{Henning}, J.~W., {Sayre}, J.~T., {Reichardt}, C.~L., {et~al.} 2018, \apj, 852, 97, \dodoi{10.3847/1538-4357/aa9ff4}

\bibitem[{{Hou} {et~al.}(2018){Hou}, {Aylor}, {Benson}, {Bleem}, {Carlstrom}, {Chang}, {Cho}, {Chown}, {Crawford}, {Crites}, {de Haan}, {Dobbs}, {Everett}, {Follin}, {George}, {Halverson}, {Harrington}, {Holder}, {Holzapfel}, {Hrubes}, {Keisler}, {Knox}, {Lee}, {Leitch}, {Luong-Van}, {Marrone}, {McMahon}, {Meyer}, {Millea}, {Mocanu}, {Mohr}, {Natoli}, {Omori}, {Padin}, {Pryke}, {Reichardt}, {Ruhl}, {Sayre}, {Schaffer}, {Shirokoff}, {Staniszewski}, {Stark}, {Story}, {Vanderlinde}, {Vieira}, \& {Williamson}}]{Hou:2018}
{Hou}, Z., {Aylor}, K., {Benson}, B.~A., {et~al.} 2018, \apj, 853, 3, \dodoi{10.3847/1538-4357/aaa3ef}

\bibitem[{{Hoyt} {et~al.}(2024){Hoyt}, {Jang}, {Freedman}, {Madore}, {Lee}, \& {Owens}}]{Hoyt:2024a}
{Hoyt}, T.~J., {Jang}, I.~S., {Freedman}, W.~L., {et~al.} 2024, arXiv e-prints, arXiv:2407.07309, \dodoi{10.48550/arXiv.2407.07309}

\bibitem[{{Huang} {et~al.}(2023){Huang}, {Marengo}, {Breuval}, {Hack}, {Riess}, \& {Yuan}}]{2023jwst.prop.4087H}
{Huang}, C., {Marengo}, M., {Breuval}, L., {et~al.} 2023, {Refining the Mira Distance Ladder with NIRCam Observations of M101}, JWST Proposal. Cycle 2, ID. \#4087

\bibitem[{{Huang} {et~al.}(2024){Huang}, {Yuan}, {Riess}, {Hack}, {Whitelock}, {Zakamska}, {Casertano}, {Macri}, {Marengo}, {Menzies}, \& {Smith}}]{Huang:2024}
{Huang}, C.~D., {Yuan}, W., {Riess}, A.~G., {et~al.} 2024, \apj, 963, 83, \dodoi{10.3847/1538-4357/ad1ff8}

\bibitem[{{Jang} {et~al.}(2021){Jang}, {Hoyt}, {Beaton}, {Freedman}, {Madore}, {Lee}, {Neeley}, {Monson}, {Rich}, \& {Seibert}}]{Jang:2021}
{Jang}, I.~S., {Hoyt}, T.~J., {Beaton}, R.~L., {et~al.} 2021, \apj, 906, 125, \dodoi{10.3847/1538-4357/abc8e9}

\bibitem[{{Lee}(2023)}]{Lee:2023}
{Lee}, A.~J. 2023, \apj, 956, 15, \dodoi{10.3847/1538-4357/acee69}

\bibitem[{{Lee} {et~al.}(2024){Lee}, {Freedman}, {Madore}, {Jang}, {Owens}, \& {Hoyt}}]{Lee:2024}
{Lee}, A.~J., {Freedman}, W.~L., {Madore}, B.~F., {et~al.} 2024, arXiv e-prints, arXiv:2408.03474, \dodoi{10.48550/arXiv.2408.03474}

\bibitem[{{Li} {et~al.}(2024{\natexlab{a}}){Li}, {Riess}, {Casertano}, {Anand}, {Scolnic}, {Yuan}, {Breuval}, \& {Huang}}]{Li:2024}
{Li}, S., {Riess}, A.~G., {Casertano}, S., {et~al.} 2024{\natexlab{a}}, arXiv e-prints, arXiv:2401.04777, \dodoi{10.48550/arXiv.2401.04777}

\bibitem[{{Li} {et~al.}(2024{\natexlab{b}}){Li}, {Anand}, {Riess}, {Casertano}, {Yuan}, {Breuval}, {Macri}, {Scolnic}, {Beaton}, \& {Anderson}}]{Li:2024b}
{Li}, S., {Anand}, G.~S., {Riess}, A.~G., {et~al.} 2024{\natexlab{b}}, arXiv e-prints, arXiv:2408.00065, \dodoi{10.48550/arXiv.2408.00065}

\bibitem[{{Madore} \& {Freedman}(2020)}]{Madore:2020}
{Madore}, B.~F., \& {Freedman}, W.~L. 2020, \apj, 899, 66, \dodoi{10.3847/1538-4357/aba045}

\bibitem[{{Madore} {et~al.}(2023){Madore}, {Freedman}, \& {Owens}}]{Madore:2023}
{Madore}, B.~F., {Freedman}, W.~L., \& {Owens}, K. 2023, \aj, 166, 224, \dodoi{10.3847/1538-3881/ad022c}

\bibitem[{{McQuinn} {et~al.}(2019){McQuinn}, {Boyer}, {Skillman}, \& {Dolphin}}]{McQuinn:2019}
{McQuinn}, K. B.~W., {Boyer}, M., {Skillman}, E.~D., \& {Dolphin}, A.~E. 2019, \apj, 880, 63, \dodoi{10.3847/1538-4357/ab2627}

\bibitem[{{Newman} {et~al.}(2024){Newman}, {McQuinn}, {Skillman}, {Boyer}, {Cohen}, {Dolphin}, \& {Telford}}]{Newman:2024}
{Newman}, M. J.~B., {McQuinn}, K. B.~W., {Skillman}, E.~D., {et~al.} 2024, arXiv e-prints, arXiv:2403.03086, \dodoi{10.48550/arXiv.2403.03086}

\bibitem[{{Parada} {et~al.}(2021){Parada}, {Heyl}, {Richer}, {Ripoche}, \& {Rousseau-Nepton}}]{Parada:2021}
{Parada}, J., {Heyl}, J., {Richer}, H., {Ripoche}, P., \& {Rousseau-Nepton}, L. 2021, \mnras, 501, 933, \dodoi{10.1093/mnras/staa3750}

\bibitem[{{Reid} {et~al.}(2019){Reid}, {Pesce}, \& {Riess}}]{Reid:2019}
{Reid}, M.~J., {Pesce}, D.~W., \& {Riess}, A.~G. 2019, \apjl, 886, L27, \dodoi{10.3847/2041-8213/ab552d}

\bibitem[{{Riess} {et~al.}(2021){Riess}, {Anderson}, {Breuval}, {Casertano}, {Macri}, {Scolnic}, \& {Yuan}}]{2021jwst.prop.1685R}
{Riess}, A., {Anderson}, R.~I., {Breuval}, L., {et~al.} 2021, {Uncrowding the Cepheids for an Improved Determination of the Hubble Constant}, JWST Proposal. Cycle 1, ID. \#1685

\bibitem[{{Riess} {et~al.}(2023{\natexlab{a}}){Riess}, {Anderson}, {Breuval}, {Casertano}, {Macri}, {Scolnic}, \& {Yuan}}]{2023jwst.prop.2875R}
---. 2023{\natexlab{a}}, {Scrutinizing the Dirtiest Cepheids, a Test of the Hubble Tension}, JWST Proposal. Cycle 2, ID. \#2875

\bibitem[{{Riess} {et~al.}(2019){Riess}, {Casertano}, {Yuan}, {Macri}, \& {Scolnic}}]{Riess:2019}
{Riess}, A.~G., {Casertano}, S., {Yuan}, W., {Macri}, L.~M., \& {Scolnic}, D. 2019, \apj, 876, 85, \dodoi{10.3847/1538-4357/ab1422}

\bibitem[{{Riess} {et~al.}(2011){Riess}, {Macri}, {Casertano}, {Lampeitl}, {Ferguson}, {Filippenko}, {Jha}, {Li}, \& {Chornock}}]{Riess:2011}
{Riess}, A.~G., {Macri}, L., {Casertano}, S., {et~al.} 2011, \apj, 730, 119, \dodoi{10.1088/0004-637X/730/2/119}

\bibitem[{{Riess} {et~al.}(2016){Riess}, {Macri}, {Hoffmann}, {Scolnic}, {Casertano}, {Filippenko}, {Tucker}, {Reid}, {Jones}, {Silverman}, {Chornock}, {Challis}, {Yuan}, {Brown}, \& {Foley}}]{Riess:2016}
{Riess}, A.~G., {Macri}, L.~M., {Hoffmann}, S.~L., {et~al.} 2016, \apj, 826, 56, \dodoi{10.3847/0004-637X/826/1/56}

\bibitem[{{Riess} {et~al.}(2018){Riess}, {Casertano}, {Yuan}, {Macri}, {Bucciarelli}, {Lattanzi}, {MacKenty}, {Bowers}, {Zheng}, {Filippenko}, {Huang}, \& {Anderson}}]{Riess:2018}
{Riess}, A.~G., {Casertano}, S., {Yuan}, W., {et~al.} 2018, \apj, 861, 126, \dodoi{10.3847/1538-4357/aac82e}

\bibitem[{{Riess} {et~al.}(2022){Riess}, {Yuan}, {Macri}, {Scolnic}, {Brout}, {Casertano}, {Jones}, {Murakami}, {Anand}, {Breuval}, {Brink}, {Filippenko}, {Hoffmann}, {Jha}, {D'arcy Kenworthy}, {Mackenty}, {Stahl}, \& {Zheng}}]{Riess:2022}
{Riess}, A.~G., {Yuan}, W., {Macri}, L.~M., {et~al.} 2022, \apjl, 934, L7, \dodoi{10.3847/2041-8213/ac5c5b}

\bibitem[{{Riess} {et~al.}(2023{\natexlab{b}}){Riess}, {Anand}, {Yuan}, {Casertano}, {Dolphin}, {Macri}, {Breuval}, {Scolnic}, {Perrin}, \& {Anderson}}]{Riess:2023}
{Riess}, A.~G., {Anand}, G.~S., {Yuan}, W., {et~al.} 2023{\natexlab{b}}, \apjl, 956, L18, \dodoi{10.3847/2041-8213/acf769}

\bibitem[{{Riess} {et~al.}(2024){Riess}, {Anand}, {Yuan}, {Casertano}, {Dolphin}, {Macri}, {Breuval}, {Scolnic}, {Perrin}, \& {Anderson}}]{Riess:2024}
---. 2024, \apjl, 962, L17, \dodoi{10.3847/2041-8213/ad1ddd}

\bibitem[{{Ripoche} {et~al.}(2020){Ripoche}, {Heyl}, {Parada}, \& {Richer}}]{Ripoche:2020}
{Ripoche}, P., {Heyl}, J., {Parada}, J., \& {Richer}, H. 2020, \mnras, 495, 2858, \dodoi{10.1093/mnras/staa1346}

\bibitem[{{Scolnic} {et~al.}(2020){Scolnic}, {Smith}, {Massiah}, {Wiseman}, {Brout}, {Kessler}, {Davis}, {Foley}, {Galbany}, {Hinton}, {Hounsell}, {Kelsey}, {Lidman}, {Macaulay}, {Morgan}, {Nichol}, {M{\"o}ller}, {Popovic}, {Sako}, {Sullivan}, {Thomas}, {Tucker}, {Abbott}, {Aguena}, {Allam}, {Annis}, {Avila}, {Bechtol}, {Bertin}, {Brooks}, {Burke}, {Rosell}, {Carollo}, {Kind}, {Carretero}, {Costanzi}, {da Costa}, {De Vicente}, {Desai}, {Diehl}, {Doel}, {Drlica-Wagner}, {Eckert}, {Eifler}, {Everett}, {Flaugher}, {Fosalba}, {Frieman}, {Garc{\'\i}a-Bellido}, {Gaztanaga}, {Gerdes}, {Glazebrook}, {Gruen}, {Gruendl}, {Gschwend}, {Gutierrez}, {Hartley}, {Hollowood}, {Honscheid}, {James}, {Kuehn}, {Kuropatkin}, {Lewis}, {Li}, {Lima}, {Maia}, {Marshall}, {Menanteau}, {Miquel}, {Palmese}, {Paz-Chinch{\'o}n}, {Plazas}, {Pursiainen}, {Sanchez}, {Scarpine}, {Schubnell}, {Serrano}, {Sevilla-Noarbe}, {Sommer}, {Suchyta}, {Swanson}, {Tarle}, {Varga}, {Walker}, {Wilkinson}, \& {DES Collaboration}}]{Scolnic20}
{Scolnic}, D., {Smith}, M., {Massiah}, A., {et~al.} 2020, \apjl, 896, L13, \dodoi{10.3847/2041-8213/ab8735}

\bibitem[{{Scolnic} {et~al.}(2022){Scolnic}, {Brout}, {Carr}, {Riess}, {Davis}, {Dwomoh}, {Jones}, {Ali}, {Charvu}, {Chen}, {Peterson}, {Popovic}, {Rose}, {Wood}, {Brown}, {Chambers}, {Coulter}, {Dettman}, {Dimitriadis}, {Filippenko}, {Foley}, {Jha}, {Kilpatrick}, {Kirshner}, {Pan}, {Rest}, {Rojas-Bravo}, {Siebert}, {Stahl}, \& {Zheng}}]{Scolnic:2022}
{Scolnic}, D., {Brout}, D., {Carr}, A., {et~al.} 2022, \apj, 938, 113, \dodoi{10.3847/1538-4357/ac8b7a}

\bibitem[{{Story} {et~al.}(2013){Story}, {Reichardt}, {Hou}, {Keisler}, {Aird}, {Benson}, {Bleem}, {Carlstrom}, {Chang}, {Cho}, {Crawford}, {Crites}, {de Haan}, {Dobbs}, {Dudley}, {Follin}, {George}, {Halverson}, {Holder}, {Holzapfel}, {Hoover}, {Hrubes}, {Joy}, {Knox}, {Lee}, {Leitch}, {Lueker}, {Luong-Van}, {McMahon}, {Mehl}, {Meyer}, {Millea}, {Mohr}, {Montroy}, {Padin}, {Plagge}, {Pryke}, {Ruhl}, {Sayre}, {Schaffer}, {Shaw}, {Shirokoff}, {Spieler}, {Staniszewski}, {Stark}, {van Engelen}, {Vanderlinde}, {Vieira}, {Williamson}, \& {Zahn}}]{Story:2013}
{Story}, K.~T., {Reichardt}, C.~L., {Hou}, Z., {et~al.} 2013, \apj, 779, 86, \dodoi{10.1088/0004-637X/779/1/86}

\bibitem[{{Tully}(2023)}]{Tully:2023}
{Tully}, R.~B. 2023, arXiv e-prints, arXiv:2305.11950, \dodoi{10.48550/arXiv.2305.11950}

\bibitem[{{Uddin} {et~al.}(2023){Uddin}, {Burns}, {Phillips}, {Suntzeff}, {Freedman}, {Brown}, {Morrell}, {Hamuy}, {Krisciunas}, {Wang}, {Hsiao}, {Goobar}, {Perlmutter}, {Lu}, {Stritzinger}, {Anderson}, {Ashall}, {Hoeflich}, {Shappee}, {Persson}, {Piro}, {Baron}, {Contreras}, {Galbany}, {Kumar}, {Shahbandeh}, {Davis}, {Anais}, {Busta}, {Campillay}, {Castell{\'o}n}, {Corco}, {Diamond}, {Gall}, {Gonzalez}, {Holmbo}, {Roth}, {Ser{\'o}n}, {Taddia}, {Torres}, {Baltay}, {Folatelli}, {Hadjiyska}, {Kasliwal}, {Nugent}, {Rabinowitz}, \& {Ryder}}]{Uddin:2023}
{Uddin}, S.~A., {Burns}, C.~R., {Phillips}, M.~M., {et~al.} 2023, arXiv e-prints, arXiv:2308.01875, \dodoi{10.48550/arXiv.2308.01875}

\bibitem[{{Verde} {et~al.}(2023){Verde}, {Sch{\"o}neberg}, \& {Gil-Mar{\'\i}n}}]{Verde:2023}
{Verde}, L., {Sch{\"o}neberg}, N., \& {Gil-Mar{\'\i}n}, H. 2023, arXiv e-prints, arXiv:2311.13305, \dodoi{10.48550/arXiv.2311.13305}

\bibitem[{{Weisz} {et~al.}(2024){Weisz}, {Dolphin}, {Savino}, {McQuinn}, {Newman}, {Williams}, {Kallivayalil}, {Anderson}, {Boyer}, {Correnti}, {Geha}, {Sandstrom}, {Cole}, {Warfield}, {Skillman}, {Cohen}, {Beaton}, {Bressan}, {Bolatto}, {Boylan-Kolchin}, {Brooks}, {Bullock}, {Conroy}, {Cooper}, {Dalcanton}, {Dotter}, {Fritz}, {Garling}, {Gennaro}, {Gilbert}, {Girardi}, {Johnson}, {Johnson}, {Kalirai}, {Kirby}, {Lang}, {Marigo}, {Richstein}, {Schlafly}, {Tollerud}, \& {Wetzel}}]{2024ApJS..271...47W}
{Weisz}, D.~R., {Dolphin}, A.~E., {Savino}, A., {et~al.} 2024, \apjs, 271, 47, \dodoi{10.3847/1538-4365/ad2600}

\bibitem[{{Wu} {et~al.}(2023){Wu}, {Scolnic}, {Riess}, {Anand}, {Beaton}, {Casertano}, {Ke}, \& {Li}}]{Wu2023}
{Wu}, J., {Scolnic}, D., {Riess}, A.~G., {et~al.} 2023, \apj, 954, 87, \dodoi{10.3847/1538-4357/acdd7b}

\bibitem[{{Wu} {et~al.}(2014){Wu}, {Tully}, {Rizzi}, {Dolphin}, {Jacobs}, \& {Karachentsev}}]{2014AJ....148....7W}
{Wu}, P.-F., {Tully}, R.~B., {Rizzi}, L., {et~al.} 2014, \aj, 148, 7, \dodoi{10.1088/0004-6256/148/1/7}

\end{thebibliography}

\appendix
\section{Data tables}
Here we include all the distance measurements discussed in this paper from \cite{Riess:2022}, \cite{Freedman:2024}, \cite{Li:2024}, \cite{Li:2024b}, \cite{Freedman:2021} and others. They are presented in Tables~\ref{tab:distpar} \& \ref{tab:matrix}.

We show an example plot that can be made using the data table. NGC$\,$5643 has the largest number of measurements from different teams and techniques. We show the agreement of these results in Fig.~\ref{fg:comparedis5643}. We augment the plot with data from \cite{Tully:2023}.
\setcounter{table}{0}
\setcounter{figure}{0}
\renewcommand*\thetable{A\arabic{table}}
\renewcommand*\thefigure{A\arabic{figure}}

\begin{figure}[b!] 
\begin{center}
\includegraphics[width=\textwidth]{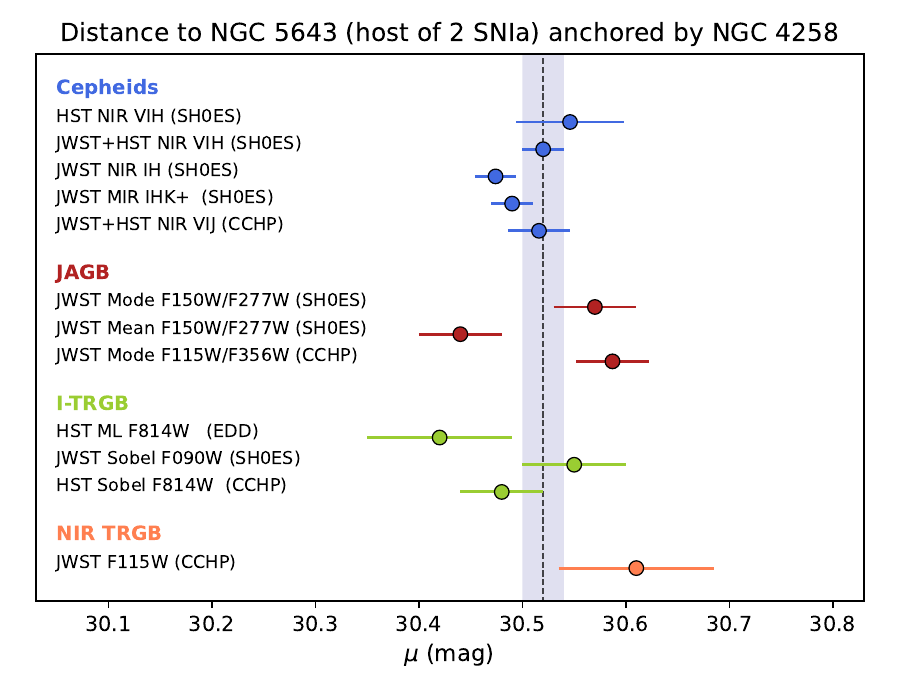}
\end{center}
\caption{Distance comparisons for NGC$\,$5643 from multiple analyses. The SH0ES measurements can be found in R22 and \cite{Riess:2024} for Cepheids, \cite{Li:2024} for JAGB, and here  for {\it JWST} TRGB. EDD measurements are from \cite{Anand:2022}. The CCHP measurements can be found in \cite{Freedman:2021} and \cite{Freedman:2024}   All measurements are anchored to NGC$\,$4258 in the first rung. We plot a consensus mean of $\mu=30.52 \pm 0.03$~mag.}
\label{fg:comparedis5643} 
\end{figure} 

\begin{deluxetable}{lrrlrrllll}
\tabletypesize{\scriptsize}
\tablewidth{0pc}
\tablecaption{Mean SN~Ia $M_B^0$ derived from Subsamples Calibrated by {\it HST} Cepheids, NGC$\,$4258, and Pantheon$+$ \citepalias{Riess:2022} \label{tab:distpar}}
\tablehead{\colhead{Host} & \colhead{$\mu_{\rm Ceph}$} & \colhead{$\sigma$} & \colhead{SN} & \colhead{$m_{B,i}^0$} & \colhead{$\sigma$} & \colhead{$M_{B,i}^0$} & \colhead{$\sigma$} & \colhead{H$_0$} & \colhead{$\sigma$}\\[-0.3cm]
\colhead{} & \multicolumn{2}{c}{[mag]} & \colhead{} & \multicolumn{4}{c}{[mag]} & \multicolumn{2}{c}{[km/s/Mpc]}}  
\startdata
N4258 &  29.398  &  0.025$^a$  &  -- &  --  &  --  &  --  &  -- & -- & --  \\
\hline
\multicolumn{10}{c}{CCHP {\it JWST} Selected Targets, $N=11$} \\
\hline
N7250 &  31.642  &  0.130  &  2013dy &  12.283  &  0.178  &  -19.357  &  0.222 &  &  \\
N4536 &  30.870  &  0.061  &  1981B &  11.551  &  0.133  &  -19.320  &  0.147 &  &     \\
N3972$^b$ &  31.644  &  0.096  &  2011by &  12.548  &  0.094  &  -19.103  &  0.136 &  &    \\
N4424$^b$ &  30.854  &  0.133  &  2012cg &  11.487  &  0.192  &  -19.376  &  0.236 &  &  \\
N4639 &  31.823  &  0.091  &  1990N &  12.454  &  0.124  &  -19.373  &  0.155 &  &    \\
N4038$^b$ &  31.612  &  0.121  &  2007sr &  12.409  &  0.106  &  -19.212  &  0.164 &  &    \\
M101  &  29.188  &  0.055  &  2011fe &  9.7800  &  0.115  &  -19.421  &  0.127 &  &    \\
N2442 &  31.459  &  0.073  &  2015F &  12.234  &  0.082  &  -19.230  &  0.111 &  &    \\
N1365 &  31.378  &  0.061  &  2012fr &  11.900  &  0.092  &  -19.478  &  0.112 &  &    \\
N5643 &  30.553  &  0.063  &  2013aa &  11.252  &  0.079  &  -19.322  &  0.101 &  &     \\
N5643 &  30.553  &  0.063  &  2017cbv &  11.208  &  0.074  &  -19.366  &  0.097 &  &    \\[0.0cm]
\hline
\multicolumn {6}{r}{Mean } & {-19.32} & 0.05 & 71.2 & 2.0 \\
\hline
\multicolumn {6}{r}{For JAGB ($N=8$): Mean } & {-19.34} & 0.05 & 70.3 & 2.0\\
\hline
\multicolumn{10}{c}{SH0ES {\it JWST} Selected/System Targets, $N=11$} \\
\hline
N1559 &  31.500  &  0.071  &  2005df &  12.141  &  0.086  &  -19.367  &  0.112  &  &    \\
N1448 &  31.298  &  0.051  &  2001el &  12.254  &  0.136  &  -19.046  &  0.146  &  &    \\
N1448 &  31.298  &  0.051  &  2021pit &  11.752  &  0.200  &  -19.548  &  0.207 &  &    \\
N5584 &  31.766  &  0.062  &  2007af &  12.804  &  0.079  &  -18.963  &  0.102  &  &    \\
N5643 &  30.553  &  0.063  &  2013aa &  11.252  &  0.079  &  -19.322  &  0.101 &  &     \\
N5643 &  30.553  &  0.063  &  2017cbv &  11.208  &  0.074  &  -19.366  &  0.097 &  &    \\
N5468 &  33.127  &  0.082  &  1999cp &  13.880  &  0.080  &  -19.248  &  0.116 &  &     \\
N5468 &  33.127  &  0.082  &  2002cr &  13.993  &  0.072  &  -19.135  &  0.111 &  &    \\
N4038$^b$ &  31.612  &  0.121  &  2007sr &  12.409  &  0.106  &  -19.212  &  0.164 &  &    \\
N3447 &  31.947  &  0.049  &  2012ht &  12.736  &  0.089  &  -19.213  &  0.102 &  &    \\
M101$^c$ &  29.188  &  0.055  &  2011fe &  9.7800  &  0.115  &  -19.421  &  0.127 &  &    \\[-0.0cm]
\hline
\multicolumn {6}{r}{ Mean } & {-19.25} & 0.05 & 73.9 & 2.0 \\
\hline
\multicolumn {6}{r}{ {\bf Mean all {\it JWST} $D\leq 25$ Mpc, 16 SNe~Ia}} & {-19.27} & {0.04} & 72.8 & 1.8 \\
\hline
\hline
\multicolumn{10}{c}{SH0ES {\it JWST} Cycle 2 observed} \\
\hline
N2525 &  32.059  &  0.105  &  2018gv &  12.728  &  0.074  &  -19.344  &  0.130  &  &    \\
N3370 &  32.132  &  0.062  &  1994ae &  12.937  &  0.082  &  -19.196  &  0.104 &  &    \\
N5861 &  32.232  &  0.105  &  2017erp &  12.945  &  0.107  &  -19.294  &  0.152 &  &    \\
N3147 &  33.173  &  0.163  &  2021hpr &  13.843  &  0.159  &  -19.358  &  0.230  &  &   \\
N3147 &  33.173  &  0.163  &  1997bq &  13.821  &  0.141  &  -19.380  &  0.218 &  &     \\
N3147 &  33.173  &  0.163  &  2008fv &  13.936  &  0.200  &  -19.264  &  0.260 &  &    \\[-0.0cm]
\hline
\multicolumn {6}{r}{{\bf Mean all {\it JWST} observed, 24 SNe~Ia}} & {-19.26} & {0.03} & 73.1 & 1.6 \\
\hline
\multicolumn{10}{c}{Remaining {\it HST} Subsample} \\
\hline
N3021 &  32.473  &  0.162  &  1995al &  13.114  &  0.116  &  -19.368  &  0.203 &  &    \\
N1309 &  32.552  &  0.069  &  2002fk &  13.209  &  0.082  &  -19.345  &  0.108 &  &    \\
N3982 &  31.736  &  0.080  &  1998aq &  12.252  &  0.078  &  -19.484  &  0.113 &  &     \\
N1015 &  32.691  &  0.077  &  2009ig &  13.350  &  0.094  &  -19.346  &  0.123 &  &    \\
N5917 &  32.377  &  0.125  &  2005cf &  13.079  &  0.095  &  -19.297  &  0.160 &  &    \\
U9391 &  32.861  &  0.075  &  2003du &  13.525  &  0.084  &  -19.335  &  0.114 &  &     \\
N3583 &  32.814  &  0.087  &  ASASSN-15so &  13.509  &  0.093  &  -19.308 &  0.129 & & \\
N2608 &  32.620  &  0.158  &  2001bg &  13.443  &  0.166  &  -19.191  &  0.232 &  &    \\
N7541 &  32.512  &  0.124  &  1998dh &  13.418  &  0.128  &  -19.095  &  0.181 &  &   \\
N0691 &  32.838  &  0.114  &  2005W &  13.602  &  0.139  &  -19.250  &  0.182  &  &   \\
N3254 &  32.343  &  0.084  &  2019np &  13.201  &  0.074  &  -19.141  &  0.114 &  &     \\
N5728 &  33.101  &  0.208  &  2009Y &  13.514  &  0.115  &  -19.607  &  0.242  &  &   \\
N7678 &  33.196  &  0.157  &  2002dp &  14.090  &  0.093  &  -19.113  &  0.187 &  &    \\
M1337 &  33.060  &  0.121  &  2006D &  13.655  &  0.106  &  -19.406  &  0.164 &  &    \\
N4680 &  32.609  &  0.208  &  1997bp &  13.173  &  0.205  &  -19.440  &  0.296 &  &     \\
N7329 &  33.252  &  0.122  &  2006bh &  14.030  &  0.079  &  -19.248  &  0.146 &  &    \\
N0976 &  33.719  &  0.153  &  1999dq &  14.250  &  0.103  &  -19.475  &  0.188 &  &    \\
N0105 &  34.538  &  0.252  &  2007A &  15.250  &  0.133  &  -19.290  &  0.292 &  &     \\[0.0cm]
\hline
\multicolumn {6}{c}{Mean of all, unique 42 SNe~Ia } & {-19.28} & {0.03} & 72.5 & 1.5 \\
\hline
\enddata
\tablecomments{(a) Includes intercept uncertainty from Cepheid sample and tie to NGC$\,$4258 sample with different mean period; see Table~\ref{table:h0err2}. (b) Hosts excluded by CCHP JAGB analysis. (c) {\it JWST} program GO-2581. (d) {\it JWST} program GO-4087. Sample errors in mean include measurement errors for NGC$\,$4258 but not geometric distance error of 0.032 mag. H$_0$ follows from $<M_B^0>$ with Pantheon+ SNe as 5\,log(H$_0/\MBh)=M_B^0-(\MB)$.} 
\end{deluxetable}

\begin{sidewaystable}
    \caption{Distance Moduli for Comparison\label{tab:matrix}}
    \begin{center}
    \begin{tabular}{lllllllllllllllllll}
    \hline
    & \multicolumn{2}{c}{HST Cepheids} & \multicolumn{4}{c}{JWST Cepheids} & \multicolumn{4}{c}{JWST TRGB} & \multicolumn{2}{c}{HST TRGB} & \multicolumn{4}{c}{JWST JAGB} & \multicolumn{2}{c}{Miras}\\
    \hline
    Host & R22 & err & CCHP & err & SH0ES & err & CCHP & err & SH0ES & err & F21 & err & CCHP & err & SH0ES & err & H24 & err \\
    \hline
    \hline
    N4258 & 29.4 & 0.025 & 29.4 & 0.087 & 29.4 & 0.03 & 29.4 & 0.035 & 29.4 & 0.05 & 29.4 & 0.04 & 29.4 & 0.05 & 29.4 & 0.05 & 29.4 & 0.04 \\
    \hline
    M101 & 29.188 & 0.055 & 29.14 & 0.08 & 29.12 & 0.03  & 29.18 & 0.04 & - & - & 29.08 & 0.04 & 29.22 & 0.04 & - & - & 29.1 & 0.06 \\
    N1309 & 32.552 & 0.069 & - & - & - & - & - & - & - & - & 32.49 & 0.07$^b$ & - & - & - & - & - & - \\
    N1365 & 31.378 & 0.061 & 31.26 & 0.1 & - & - & 31.33 & 0.07 & - & - & 31.36 & 0.05 & 31.39 & 0.04 & - & - & - & - \\
    N1448 & 31.298$^*$ & 0.051 & - & - & 31.289 & 0.03 & - & - & 31.38 & 0.07 & 31.32 & 0.06 & - & - & 31.29 & 0.04 & - & - \\
    N1559 & 31.500$^*$ & 0.071 & - & - & 31.371 & 0.03 & - & - & 31.5 & 0.05 & - & - & - & - & 31.39 & 0.04 & 31.41 & 0.08 \\
    N2442 & 31.459 & 0.073 & 31.47 & 0.09 & - & - & 31.61 & 0.09 & - & - & - & - & 31.61 & 0.04 & - & - & - & - \\
    N2525 & 32.059 & 0.11 & - & - & - & - & - & - & 31.81 & 0.08 & - & - & - & - & - & - & - & - \\
    N3021 & 32.473 & 0.162 & - & - & - & - & - & - & - & - & 32.22$^b$ & 0.05 & - & - & - & - & - & - \\
    N3370 & 32.130 & 0.06 & - & - & - & - & - & - & - & - & 32.27$^b$ & 0.05 & - & - & - & - & - & - \\
    N3447 & 31.947 & 0.05 & - & - & 31.95 & 0.03 & - & - & 31.92 & 0.09 & - & - & - & - & 31.85 & 0.07 & - & - \\
    N3972  & 31.644 & 0.096 & 31.67 & 0.1 & - & - & 31.74 & 0.07 & - & - & - & - & - & - & - & - & - & - \\
    N4038  & 31.612 & 0.121 & 31.7 & 0.12 & 31.67 & 0.035 & 31.61 & 0.08 & - & - & 31.68 & 0.05 & - & - & - & - & - & - \\
    N4424 & 30.854 & 0.133 & 30.91 & 0.22 & - & - & 30.93 & 0.05 & - & - & 31.0 & 0.06 & - & - & - & - & - & - \\
    N4536 & 30.870 & 0.061 & 30.95 & 0.12 & - & - & 30.94 & 0.06 & - & - & 30.96 & 0.05 & 30.98 & 0.03 & - & - & - & - \\
    N4639 & 31.823 & 0.091 & 31.8 & 0.12 & - & - & 31.75 & 0.07 & - & - & - & - & 31.74 & 0.04 & - & - & - & - \\
    N5468 & 33.127$^*$ & 0.082 & - & - & 32.975 & 0.03 & - & - & - & - & - & - & - & - & - & - & - & - \\
    N5584 & 31.766$^*$ & 0.062 & - & - & 31.838 & 0.03 & - & - & 31.81$^b$ & 0.09 & 31.82 & 0.1 & - & - & 31.85 & 0.04 & - & - \\
    N5643 & 30.553$^*$ & 0.063 & 30.51 & 0.08 & 30.52 & 0.03 & 30.61 & 0.07 & 30.56 & 0.06 & 30.475 & 0.08 & 30.59 & 0.04 & 30.49 & 0.04 & - & - \\
    N7250  & 31.642 & 0.13 & 31.41 & 0.12 & - & - & 31.62 & 0.04 & - & - & - & - & 31.6 & 0.08 & - & - & - & - \\
    \hline
    \end{tabular}\\\tiny \ \\
    \end{center}
    \small{$^*$R24 Table 3 refit R22 {\it HST} Cepheids to a common $P$--$L$ slope with {\it JWST} at the same wavelength to negate common error.  These {\it HST} distance-modulus values (mag) improve the Cepheid comparison with {\it JWST} and are N5643, $30.518 \pm 0.033$; N5584, $31.828\pm0.037$; N1559, $31.473\pm0.045$; N1448, $31.236\pm0.034$; N5468, $33.058\pm0.052$.   $^a$N2525 qualifies for the $D<25$ Mpc TRGB sample based on its distance. The uncertainties for the CCHP JWST measures in NGC 4258 were derived from table 5 in F24 after removing the 1.5\% geometric distance uncertainty. $^b$ The HST TRGB distances given in F21 for NGC 1309, 3021, 3370, and 5584, all at the far end of the measurable range, are contentious as \cite{Anand:2022} have reanalyzed them and could not detect the TRGB.  We include them here to keep the F21 sample complete.}
\end{sidewaystable}

\clearpage
\setcounter{table}{0}
\setcounter{figure}{0}
\renewcommand*\thetable{B\arabic{table}}
\renewcommand*\thefigure{B\arabic{figure}}
\section{Differences with Analyses in F24}
Here we discuss a few points which we believe may be in error in the F24 (as initially posted to the arxiv) analysis based on the considerations below.
We fully recognize that these issues may evolve with additional work.

\begin{enumerate}
    \item \textit{Method for correlated errors and $H_0$:} \\
       There are two types of uncertainties in the determination of $H_0$ which are correlated across distance methods (i.e., the same error) and so not reduced by combining distance measuring methods: the SNe uncertainty (due to the $\sim$10 SN in the hosts, and the $\sim$300 which measure the Hubble flow) and the 1.5\% uncertainty geometric distance of NGC$\,$4258. Table 5 in F24 presents random and systematic errors for each of 3 methods before fully reducing their combination. However, the random errors given in F24 Table 3 for each method include the same $\sim$10 SN Ia (and Hubble flow SNe) and the dominant error is the fit SN intrinsic scatter of 0.19 mag (Table 3 in F24) divided by $\sqrt{10-1}$, 2.9\% in $H_0$ and is the same uncertainty for all distance methods. F24 reduces this by $\sim\sqrt{3}$ as though independent. This also occurs in Figure 11 of F24 where the PDFs for the three method results, including the SNe in common, are combined by multiplying them together.
       
       A similar situation occurs in the reduction of the systematic errors for each method.   F24 gives these in their Table 5 as the sum of the irreducible 1.5\% geometric error for NGC$\,$4258 and the individual uncertainties in measuring each method in NGC$\,$4258 (e.g., TRGB fitting, smoothing parameters, Cepheid \PL cutoff, etc). Here too the method combination fully reduces these errors which also reduces the geometric distance error though it is the same for all methods. To illustrate this we separate the reducible and irreducible error terms for the F24 $H_0$ error budget in Table 7 and recalculate the error in $H_0$. We find a method combined error of 3.3\% or $\pm$ 2.3 \kmsmpc\ which matches the independent calculation of the expected uncertainty in Table 3. It is higher than the combined error of $\pm$ 1.5 \kmsmpc\ from the method combination in F24 which is a consequence of neglecting the method-correlated errors in their reduction.  

     \item \textit{Method for measurement Uncertainties in NGC$\,$4258 and Distance Comparisons:} \\
     In the point above, we separated the measurement uncertainties for each method in NGC$\,$4258 from the 1.5\% geometric distance uncertainty as provided in F24 (see Table 5, 3rd column). In magnitudes these are 0.035, 0.05, 0.087 for TRGB, JAGB and Cepheids, respectively. As given in \S 2.2, the measurement uncertainty of each method in NGC$\,$4258 is the largest component of the total comparison uncertainty because they do not average down with more SN hosts. For example, the 0.087 mag uncertainty for measuring Cepheids with {\it JWST} in NGC$\,$4258 alone is already larger than the size of the JAGB-Cepheid difference. Likewise, the mean difference between the CCHP JAGB distances (L24) and those from CCHP {\it HST} TRGB \citep{Freedman:2021} of 0.073 mag would be even more significant than JAGB-Cepheids, but not so including the NGC$\,$4258 measurement errors. In Table 8 we use the data in F24 (Table 2 and 5) to calculate the weighted mean differences and their uncertainties without these terms (following F24) and with them. As shown, all of the distance comparisons are in good accord. We conclude that the difference between the mean CCHP {\it JWST} JAGB and CCHP {\it JWST} Cepheid distances are not significant after including the neglected NGC$\,$4258 method measurement errors.

     \item \textit{Linearity of Cepheid Distances:} \\
     F24 claimed a 3$\sigma$ correlation or Cepheid-distance non-linearity by comparing {\it HST} Cepheid distances from R22, $\mu_{\it HST}$, and SN Ia absolute luminosity, $M_B$. However, the values of $M^0_B$ (also from R22) were formed from $m^0_B-\mu_{\it HST}$ so that the measured quantity, $\mu_{\it HST}$, appears on {\it both sides} of the regression (axes), and thus are intrinsically correlated.  In the {\it uncorrelated} space of {\it HST} Cepheid distance vs SN Ia $m_B$ we calculate a slope ranging from 0.032 to 0.039 $\pm0.022$ or 1.5$\sigma$ to 1.7$\sigma$ significance, depending on the regression method used, and thus we do not detect a significant dependence. Likewise, we find the difference for the near and far SN mean $M_B$ in Figure 16 of F24 to be 0.04$\pm0.05$ mag(0.9$\sigma$) and 0.07$\pm0.11$ mag(0.6$\sigma$) for the high and low SNR split, respectively, and thus not significantly different. A stronger test of HST Cepheid distances is given in \S 2.3.1 using all independent measures produces a relation of (1-slope) which is -0.006 $\pm$ 0.01. 
     
\end{enumerate}
\begin{deluxetable*}{lcccc}
\tablecaption{Reducible and irreducible H$_0$ Uncertainties from F24 \label{tab:hoerrors}} 
\tablehead{\multicolumn{1}{l}{Source of Error\hspace*{0.5in}} &\colhead{SN Mean$^f$} &\colhead{\hspace*{0.25in}Hosts Mean\hspace*{0.25in}} &\colhead{\hspace*{0.25in}N4258$^g$\hspace*{0.25in}} & \colhead{N4258}\\[-6pt]
& \colhead{Error} & \colhead{Measures} & \colhead{Measures} & \colhead{Geom. D}}
\startdata
TRGB zero point     &2.6\% &1.0\%& 1.6\%$^a$\ & 1.5\%$^e$ \\
JAGB zero point     & 2.6\% &1.0\% & 2.4\%$^b$\ & 1.5\%\\
Cepheid zero point  & 2.6\% &1.3\% & 4.0\%$^c$\ & 1.5\%\\
\hline  \hline
Combined methods$^d$& 2.6\% & 0.6\% & 1.3\% & 1.5\% \\
Combined & \multicolumn{4}{c}{\ \ \,3.3\% or 2.3 \kmsmpc } \\
\enddata
\tablecomments{(a) as per F24: NGC$\,$4258 uncertainty in color term, TRGB fitting, extinction, photometry calibration. (b) as per F24: NGC$\,$4258  Uncertainty in the mode, smoothing parameter, convergence error, extinction. (c) as per F24: Uncertainty in NGC$\,$4258 reddening law fit zero-point, NGC$\,$4258 PL cutoff, aperture correction uncertainty, including cross matching the {\it HST} catalogs, photometric zero point uncertainty and metallicity. (d) Cepheid, TRGB and JAGB combined errors, first and last columns do not reduce because they use the same SNe and NGC$\,$4258 geometric distance. (e) Uncertainty of 1.5\% in NGC$\,$4258 distance \citep{Reid:2019} (f) error from 10 SN Ia with intrinsic scatter 0.19 mag from Table 3 in F24 and error in Hubble flow SNe, host distance measures were removed from the random error in F24 Table 5 and put in column 2 (g) Errors in each methods measurement in NGC$\,$4258 after removing the geometric distance error now placed in column 4.}
\end{deluxetable*}
\begin{deluxetable*}{lcccc}
\tablecaption{CCHP Distance Method Comparisons without and with Method Measurement Errors in NGC$\,$4258\label{tab:distcompf24}} 
\tablehead{ \colhead{Comparison} & \colhead{N} & \colhead{weighted diff$^a$ (mag)} & \colhead{error} & \colhead{w/ NGC$\,$4258 measurement errors}  } 
\startdata
CCHP: {\it JWST} JAGB-JWST Cepheids     & 7 & 0.086 & $\pm 0.040$ (2.2$\sigma$) &  $\pm$ 0.108(0.8$\sigma$) \\
CCHP: {\it JWST} JAGB-JWST TRGB     & 10 & 0.019 & $\pm 0.029$ (0.7$\sigma$) & $\pm$ 0.067(0.8$\sigma$)  \\
CCHP: {\it JWST} Cepheids-JWST TRGB  & 10 & -0.059 & $\pm 0.039$(1.5$\sigma$)  & $\pm$ 0.101(0.3$\sigma$) \\
CCHP: {\it JWST} JAGB-HST TRGB(F21)     & 4 & 0.073 & $\pm 0.032$(2.3$\sigma$) & $\pm$ 0.072(1.0$\sigma$)  \\
\hline 
\enddata
\tablecomments{(a) We note some numerical differences between the weighted mean error given in F24 and what we calculate from their data Table 2. Most significant is JAGB vs Cepheids where F24 gives 0.028 and we find 0.040 which reduces the significance of the claimed difference from 3$\sigma$ claimed by F24 to 2$\sigma$ and then 0.8$\sigma$ including the measurement errors for each method in NGC$\,$4258 in Table 5 here and Table 5 in F24. }
\end{deluxetable*}
\vspace*{3.25in}
\end{document}